\documentclass[11pt]{article}
\usepackage[pdftex]{graphicx}

\usepackage{cite}
\usepackage{color}
\usepackage{amsfonts}
\usepackage{amsmath}
\usepackage{latexsym}
\usepackage{fancybox}
\usepackage{subfigure}
\usepackage{amssymb}
\usepackage{cite}
\usepackage{here}
\usepackage{mathtools}
\numberwithin{equation}{section}
\allowdisplaybreaks

\def\spa#1{\phantom{\fbox{\rule[-#1cm]{0cm}{0cm}}}}

\def\[#1\]{\begin{align}#1\end{align}}

\begin{document}

\hfuzz=100pt
\title{{\Large \bf{
Tensor network approach to 
2d Lorentzian\\ quantum Regge calculus
}}}
\author{
Yoshiyasu Ito$^{a}$\footnote{ito@eken.phys.nagoya-u.ac.jp},
Daisuke Kadoh$^{b}$\footnote{dkadoh@mail.doshisha.ac.jp}, 
Yuki Sato$^{ac}$\footnote{sato@tokuyama.ac.jp, ysato@th.phys.nagoya-u.ac.jp}
  \spa{0.6} \\
\\
$^a${\small{\it Department of Physics, Nagoya University}}
\\ {\small{\it Chikusaku, Nagoya 464-8602, Japan}}\\
\\
$^b${\small{\it Faculty of Sciences and Engineering, Doshisha University}}
\\ {\small{\it Kyoto 610-0394, Japan}}\\
\\
$^c${\small{\it Liberal Arts Division, National Institute of Technology, Tokuyama College}}
\\ {\small{\it Gakuendai, Shunan, Yamaguchi 745-8585, Japan}}
\spa{0.3} 
}
\date{}

\maketitle
\centerline{}

\begin{abstract} 
We demonstrate a  tensor renormalization group (TRG) calculation for a two-dimensional Lorentzian model of quantum Regge calculus (QRC). This model is expressed in terms of a tensor network by discretizing the continuous edge lengths of simplicial manifolds and identifying them as tensor indices. 
The expectation value of space-time area, which is obtained through the higher-order TRG method, 
nicely reproduces the exact value. The Lorentzian model does not have the spike configuration that was an obstacle in the Euclidean QRC, but it still has a length-divergent configuration called a pinched geometry. 
We find a possibility that the pinched geometry is suppressed by checking the average edge length squared in the limit where the number of simplices is large. 
This implies that the Lorentzian model may describe smooth geometries, although the investigation of the higher moments is required to make the statement more conclusive.  
Our results also indicate that TRG is a promising approach to numerical study of simplicial quantum gravity.

\end{abstract}

\renewcommand{\thefootnote}{\arabic{footnote}}
\setcounter{footnote}{0}

\newpage

\section{Introduction}
\label{sec:Introduction}
Regge calculus (RC) \cite{Regge:1961px} is an insightful approach to discretize space-time manifolds by simplices, aiming at dealing with the dynamics of space-time. 
Its application to quantum regime knowns as quantum Regge calculus (QRC) has also been extensively studied (see e.g. \cite{Williams:1991cd, Williams:1996jb, Hamber:2007fk, Barrett:2018ybl} for reviews and references therein).  
QRC uses the edge lengths of each simplex as dynamical variables\footnote{
Yet another formulation of quantum space-time based on Regge's thought is dynamical triangulations (DT) \cite{Ambjorn:1985az,Ambjorn:1985dn,David:1984tx,Billoire:1985ur,Kazakov:1985ea,Boulatov:1986jd} (see \cite{Ambjorn:1997di} for a pedagogical textbook), in which all edge lengths are kept fixed although the incidence matrices, i.e. ``triangulations,'' are dynamical. 
Causal dynamical triangulations is a Lorentzian version of DT, which is known as CDT (see \cite{Loll:2019rdj, Ambjorn:2022btk} for recent reviews). 
}. 
The average edge length which is a sort of dynamical lattice spacing should be much smaller than the characteristic curvature scale 
to get a sensible smooth geometry.  
Therefore, a \textit{finite} average edge length is a key to extract meaningful physical results out of the QRC formalism.

The two-dimensional QRC with the Euclidean signature, however, suffers from the very existence of the so-called spikes 
which are obstacles to obtain smooth geometries \cite{Ambjorn:1997ub, Rolf:1998ja}. 
The spike is a portion of geometry that can be elongated forever with the area staying small, and its existence
is essentially characterized by a divergent average edge length. 
Introducing the higher-derivative interactions, various numerical simulations of the $2$d Euclidean QRC with or without coupling to matter have been performed, e.g. \cite{Gross:1990fq, Bock:1994mq, Nishimura:1994qg, Holm:1994sy, Bittner:2000bf}.
However, without the higher derivative terms, numerical analysis does not work well due to the existence of spikes.

On the other hand, $2$d Lorentzian QRC models do not have spike configurations \cite{Tate:2011ct, Jia:2021deb}.  
If the spikes would be absent in higher dimensions as well, the Lorentzian models may end up with reasonable models of quantum gravity. 
However, the $2$d Lorentzian QRC still has a
worrisome length-divergent configuration called a pinched geometry.  
Unlike the spike, many distant triangles are required to form the pinched geometry.    
Therefore we need to explore if the Lorentzian models can generate smooth geometries or not, 
checking the dominance of the pinched geometry when the number of triangles is large. 
The Lorentzian models, however, generally have the sign problem if one wishes to study them based on the conventional Monte Carlo methods.
This is a troublesome situation when investigating the Lorentzian QRC models in a numerical manner\footnote{
Recently a proposal for simulating Lorentzian QRC models (precisely speaking, complex generalizations of QRC) by applying the ``generalized thimble algorithm'' has been proposed \cite{Jia:2021xeh}.
}.

The tensor renormalization group (TRG) method is a promising approach to numerical studies of theories with the sign problem. 
Any statistical treatments are not required in the method, i.e. the sign problem is not a problem.   
The TRG was originally introduced by Levin and Nave \cite{Levin:2006jai} in the two-dimensional classical Ising model,
and later a higher dimensional algorithm was proposed in Ref.\cite{Xie:2012}. 
In the last decade, it has been actively applied to the study of lattice field theories, 
e.g. see references of \cite{Kadoh:2022loj}; 
for the TRG application to spin foam models and a lattice formulation of gravity in a first-order formalism, see Ref.\cite{Dittrich:2014mxa} and Ref.\cite{Asaduzzaman:2019mtx}, respectively.

In this paper, we demonstrate that the TRG calculations for a $2$d Lorentzian model of QRC work well. 
We introduce a tensor network representation of the Lorentzian QRC discretizing the edge length integration
by the Gaussian quadrature as done in \cite{Kadoh:2018hqq, Kadoh:2018tis, Kadoh:2019ube}
for scalar field theories, \footnote{
An improvement in applying the Gaussian quadrature to theories with continuous variables is described in \cite{Kadoh:2018ele}.
The energy eigenvalues of supersymemtric quantum mechanics were  obtained precisely from the transfer matrix with the Gaussian quadrature.
}
and use in particular the higher-order TRG (HOTRG) method \cite{Xie:2012} for renormalizing the tensor networks. 
Our numerical studies on the average edge lengths squared indicate that the contribution of the pinched geometries might be suppressed in the limit where the number of triangles is large, although the investigation of higher moments is necessary to make the statement more conclusive. 

The rest of this article is organized as follows. 
In Section \ref{sec:QRC}, a brief introduction to the Lorentzian QRC is presented. 
The spikes in the Euclidean QRC and the pinched geometry are reviewed.  
In Section \ref{sec:TNR}, we begin with performing an analytic continuation of the Lorentzian QRC with an IR regulator, 
and derive its tensor network representation. Several exact relations are also given. 
Numerical results are presented in Section \ref{sec:NumericalResults}. 
Section \ref{sec:Discussions} is devoted to discussions.      
 
\section{Quantum Regge Calculus in two dimensions}
\label{sec:QRC}
We give an overview of a two-dimensional Lorentzian model of 
quantum Regge calculus (a $2$d Lorentzian QRC) \cite{Tate:2011ct}. 
The conventional Euclidean model has spike configurations whose existence is an obstacle for obtaining smooth geometries.  
We will see that the Lorentzian model does not have the spike configuration, 
but a certain length-divergent configuration called a pinched geometry still exists. 
The issue of pinched geometry will be investigated numerically in a later section.

\subsection{A Lorentzian model}
\label{sec:2dLQRC}

\begin{figure}[h]
\centering
\includegraphics[width=5in]{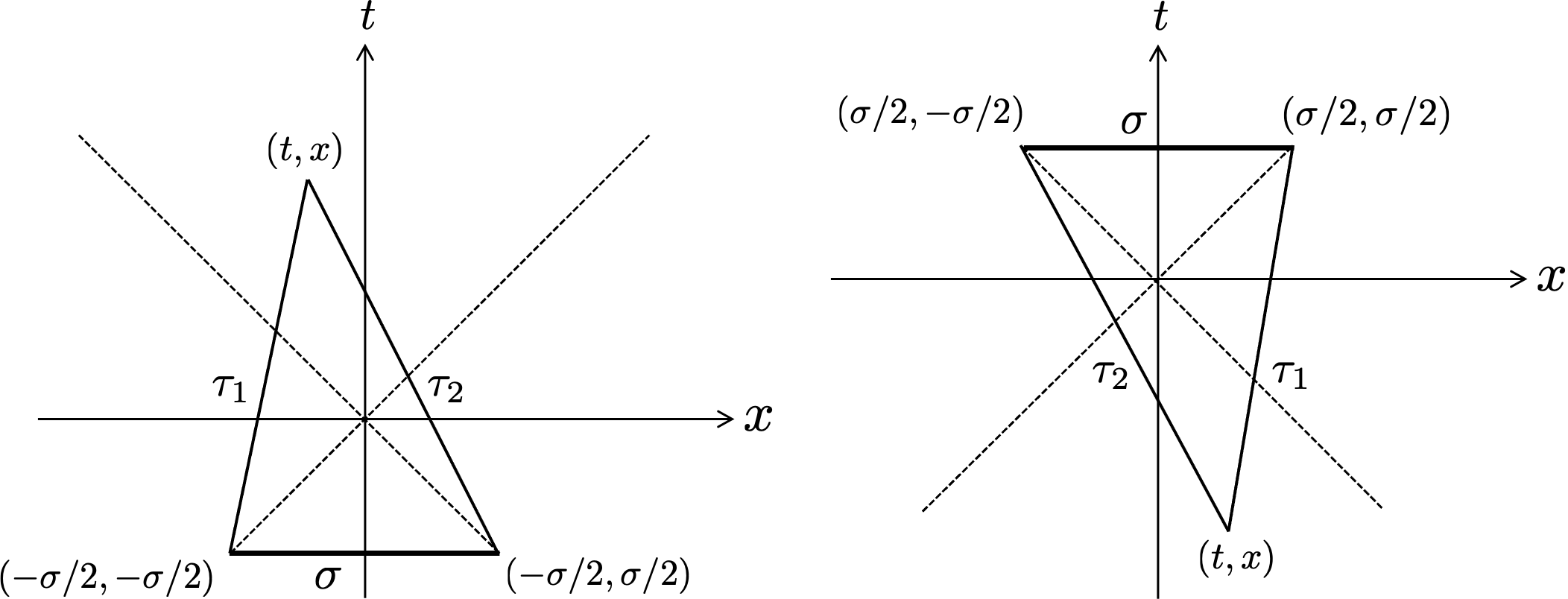}
\caption{2d Lorentzian triangles. 
The dashed lines represent the light rays. 
The thin lines ($\tau_1,\tau_2$) 
and the thick lines ($\sigma$) are the time-like edges and the space-like edges, respectively.}
\label{fig:Lorentziant}
\end{figure}

Regge calculus discretizes a space-time manifold by simplicial decompositions without introducing coordinates. 
Unlike dynamical triangulations, 
the dynamical variables in Regge calculus are the edge lengths in a fixed triangulation. 
In the $2$d Lorentzian model, building blocks are two kinds of Lorentzian triangle 
defined in the Minkowski space-time as shown in Fig.~\ref{fig:Lorentziant}. 
Each triangle has a space-like edge length $\sigma$ and two time-like edge lengths $\tau_1,\tau_2$.

In the left figure of Fig.~\ref{fig:Lorentziant}, an upward triangle is made of three vertices $(-\sigma/2,-\sigma/2)$, $(-\sigma/2,\sigma/2)$ and $(t,x)$. 
The two vertices, $(-\sigma/2,-\sigma/2)$ and $(-\sigma/2,\sigma/2)$, are placed on the past light cone 
with the proper distance $\sigma>0$ so that these vertices (events) are spatially separated. 
The third vertex $(t,x)$ lies within the future light cone as long as  $t \ge |x|$ for $t>0$. 
This vertex and the two other vertices have time-like (or null) separations characterized by the proper times $\tau_1$ and $\tau_2$:
\[
\begin{split}
&\tau^2_1 = t^2 -x^2 + \sigma (t-x) \ge 0\ , \\  
&\tau^2_2 = t^2 -x^2 + \sigma (t+x) \ge 0\ .  
\end{split}
\label{eq:tau}
\]
The same set of equations is obtained for the right figure of Fig.~\ref{fig:Lorentziant}
where $(t,x)$ lies within the past light cone.

For given edge lengths, $\sigma>0$ and $\tau_1,\tau_2 \ge 0$, a Lorentzian triangle is always created
because the third vertex $(t,x)$ can be given by 
\[
\begin{split}
& t= \pm \frac{1}{2\sigma} \left( \sqrt{ ( \sigma^2 +(\tau_1 - \tau_2)^2 )( \sigma^2 + (\tau_1 + \tau_2)^2 ) } - \sigma^2 \right)\ , \\ 
& x= \frac{1}{2\sigma} \left( \tau^2_2 - \tau^2_1 \right)\ ,
\end{split}
\label{eq:tx}  
\]
where the double sign for $t$ corresponds to the left and  right figures, respectively.
It is straightforward to show that $(t,x)$ lies within the light cone, i.e. $|t| \ge |x|$.
The area of the Lorentzian triangle is 
\[
A (\tau^2_1,\tau^2_2, \sigma^2) 
= \frac{\sigma \left( \frac{\sigma}{2} + t \right)}{2} 
= \frac{1}{4} \sqrt{ ( \sigma^2 +(\tau_1 - \tau_2)^2 )( \sigma^2 + (\tau_1 + \tau_2)^2 ) }\ , 
\label{eq:area}
\]  
which is a non-negative real number for any $\sigma>0$ and $\tau_1,\tau_2 \ge 0$.
The Euclidean QRC reviewed in the next section needs the triangle inequalities for edge lengths
to create a triangle from given three edge lengths.  
In the Lorentzian signature, on the other hand, such an extra constraint is not necessary. 
The Lorentzian QRC is therefore not related to the Euclidean QRC by a naive Wick rotation 
$\tau_1,\tau_2\rightarrow i\tau_1,i\tau_2$.

\begin{figure}[h]
\centering
\includegraphics[width=3.5in]{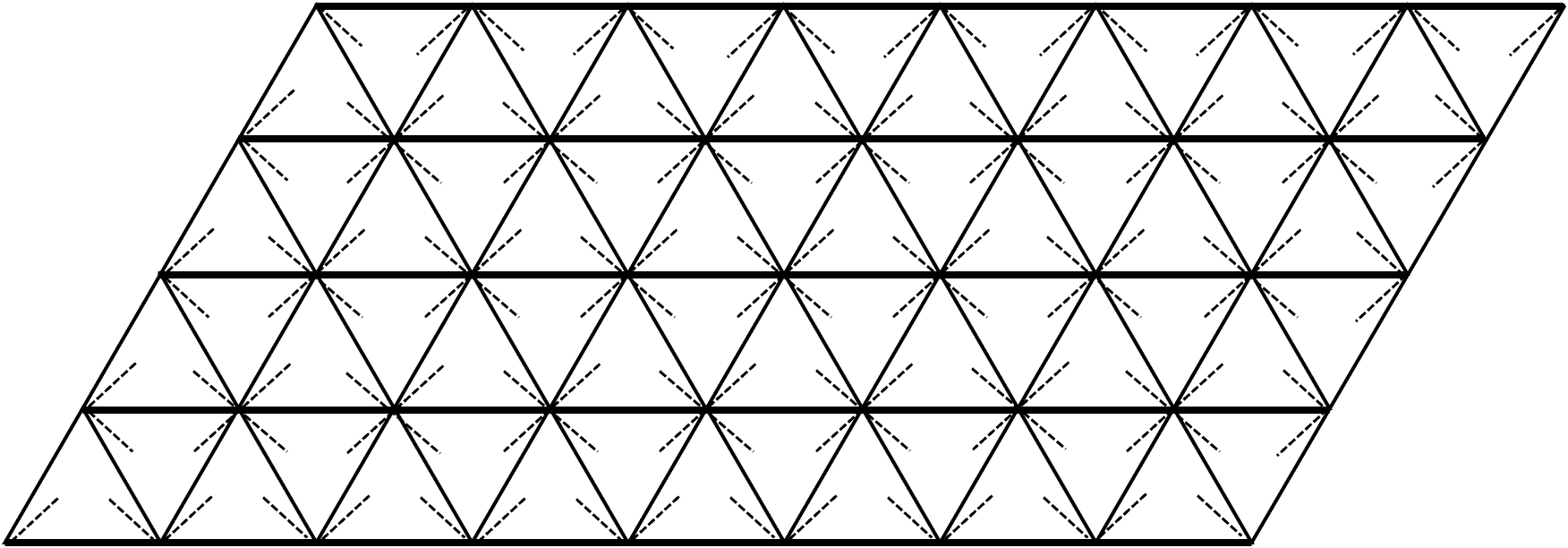}
\caption{A fixed triangulation: 
The dashed lines, the thin lines and the thick lines are the light rays, the time-like edges and the space-like edges, respectively. 
Each building block is the Lorentzian triangle, and exactly four light rays emanate from each vertex.}
\label{fig:triangulation}
\end{figure}

We consider a $2$d Lorentzian QRC with
a fixed regular triangulation shown in Fig.~\ref{fig:triangulation}, and choose the topology to be a cylinder for convenience.   
The triangulation is made of $N$ triangles where $N$ is an even integer 
since the triangulation consists of pairs of the upward and downward triangles
\footnote{Although more general triangulation can be used to define Lorentzian models of QRC, 
this triangulation is suitable for deriving a homogeneous tensor network in section \ref{sec:TNR}.}. 
The coordination number (the number of edges attached to each vertex) is fixed to $6$, 
and precisely four light rays emanate from each vertex. 
In this setup, a single light cone is defined at each vertex.  
We denote the area of the $s$-th triangle as $A_s$ for $s=1,2\cdots,N$, 
which is defined by eq.~\eqref{eq:area} as a function of the relevant three edge lengths, 
and also the $e$-th edge length as $\ell_e$ for $e=1,2,\cdots,N_e$ with the number of edges $N_e=3N/2$. 
The corresponding lattice action known as the Regge action is given by 
\[
S = \lambda \sum_{s=1}^N A_s\ , 
\label{eq:Reggeaction}
\]
with the cosmological constant $\lambda$. 
The Regge action (\ref{eq:Reggeaction}) is a discretization of the cosmological constant term $S=\displaystyle \Lambda \int d^2x \sqrt{-g}$. 

The partition function is formally given by 
\[
Z
= \int [d\ell^2]  \ 
e^{i S[\{\ell^2\}]}\ ,
\label{eq:partitionfun}
\] 
and  the expectation value of an operator ${\cal O}$ is evaluated in the standard manner as
\[
\left\langle \mathcal{O} \right\rangle 
= \frac{1}{Z} 
\int [d\ell^2] \ \mathcal{O} (\{\ell^2\}) \ 
e^{i S[\{\ell^2\}]}\ ,
\label{eq:expectationvalue}
\] 
where the integral measure is given by 
\[
[d\ell^2] = 
\prod_{e =1}^{N_e} d\ell^2_e\ \cdot f(\{ \ell^2 \})\ .
\label{eq:measureLor}
\]
Here $f(\{ \ell^2 \})$ denotes a function of the edge lengths squared. 
A few kinds of path integral measure have been discussed in previous papers because it is not uniquely determined. 
In this paper, we use in particular a local integral measure: 
\[
[d\ell^2]
= \prod_{e =1}^{N_e} d\ell^2_e\ \cdot \prod_{s=1}^N 
\left[ A_s (\{\ell^2\}) \right]^\beta\ , 
 \label{eq:measure}  
\] 
where $\beta$ is a real number specified in due course.

In Section \ref{sec:TNR}, we introduce an IR regulator to the
area function \eqref{eq:area} because it has formally a flat direction. 
Then we perform an analytic continuation from $i \lambda$ to $-|\lambda|$. 
The numerical results shown in Section  \ref{sec:NumericalResults} 
are those obtained from the analytically continued representation of the partition function.

\subsection{2d Euclidean QRC and spikes}
\label{sec:SpikesIn2dEuclideanQRC}

The $2$d Euclidean Regge calculus discretizes continuous $2$d Riemannian manifolds by triangles whose edges are straight lines in the $2$d Euclidean space. 
When fixing the topology, the Regge action is given by the summation over triangle areas: 
\[
S_E = \lambda \sum_{s =1}^N A_{E,s} \ ,
\label{eq:seuc}
\]    
where $\lambda$ is a positive cosmological constant. 
$A_{E,s}$ is the area of the $s$-th triangle, and each area is a function of the edge lengths, $\ell_1$, $\ell_2$ and $\ell_3$: 
\[
A_E = 
\frac{1}{4}
\sqrt{ - \ell^4_1 - \ell^4_2 - \ell^4_3 + 2 \left( \ell^2_1 \ell^2_2 + \ell^2_1 \ell^2_3 + \ell^2_2 \ell^2_3 \right)  }\ . 
\label{eq:ae}
\]
However, unlike the Lorentzian model, in order to create a triangle, 
all the edge lengths, $\ell_1,\ell_2,\ell_3 >0$, should satisfy the triangle inequalities:
\[
\ell_1 + \ell_2 > \ell_3\ , \ \ \ \ell_1 + \ell_3 > \ell_2\ , \ \ \ \ell_2 + \ell_3 > \ell_1\ . 
\label{eq:triangleinequalities}
\]
The partition function of the $2$d Euclidean QRC is then defined by 
\[
Z_E 
= 
\int \, [d\ell^2]_{E} \ 
e^{-  S_E(\{\ell^2\}) }\ ,
\label{eq:eucpartitionfun} 
\]
where 
\[
[d\ell^2]_{E} = \prod_{e=1}^{N_e} d\ell^2_e\ 
\Theta (\text{triangle inequalities})
\ \cdot 
f (\{\ell^2\}) 
\ . 
\label{eq:dellE}
\]
Here $f (\{\ell^2\})$ is a function of the edge length squared, 
and $\Theta (\text{triangle inequalities})$ denotes the constraint of the triangle inequalities, i.e. 
$\Theta=1$ for configurations satisfying eq.~\eqref{eq:triangleinequalities} and $\Theta=0$ otherwise. 
Note that no systematic way to determine the integral measure (\ref{eq:dellE}) is known 
although the importance of the scale-invariant measure has been pointed out in Ref.\cite{Nishimura:1994qg}. 
In addition, the integral measure is in general non-local.

It is sometimes useful to consider expectation values for a fixed area $A$: 
\[
\left\langle \mathcal{O} \right\rangle_A 
= \frac{1}{Z_{E,A}}
\int [d\ell^2]_E \ 
\mathcal{O} (\{\ell^2\}) \ 
\delta \left( A - \sum^{N}_{s=1} A_{E,s} (\{\ell^2\}) ) \right)\ , 
\label{eq:fixedaexpectationvalue}
\]
where 
\[
Z_{E,A}
= \int [d\ell^2]_E \ 
\delta \left( A - \sum^{N}_{s=1} A_{E,s} (\{\ell^2\}) ) \right)\ .
\label{eq:fixedapartitionfun}
\]
Note that $e^{-S_E}$ is a constant when the area is fixed.

\begin{figure}[h]
\centering
\includegraphics[width=3.5in]{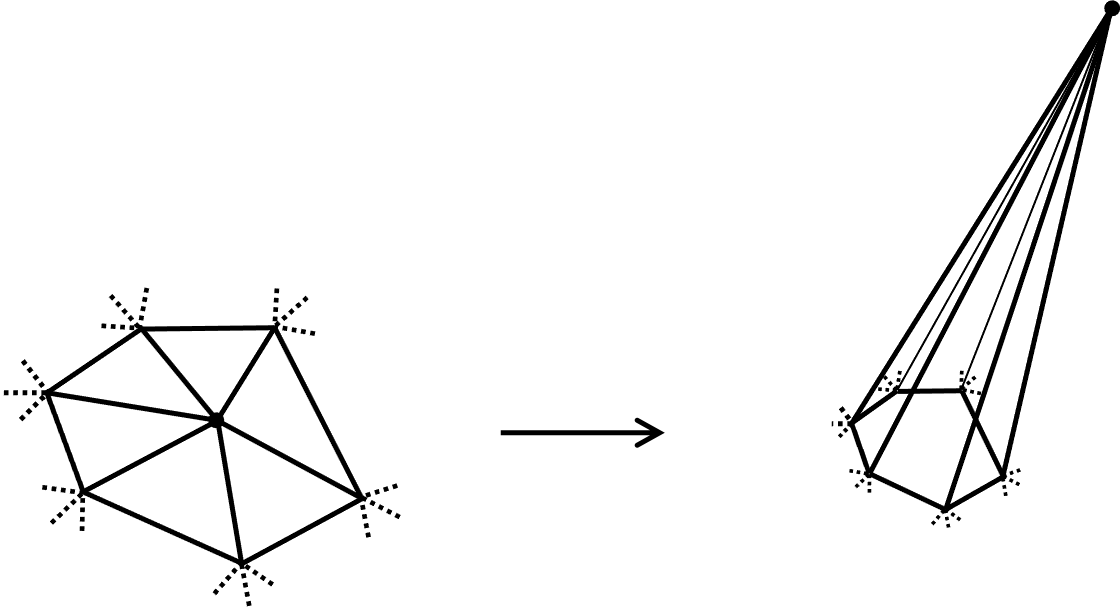}
\caption{The portion of triangulation that can be elongated without changing the area.}
\label{fig:spike}
\end{figure}

The $2$d Euclidean QRC defined above cannot be a reasonable model of quantum geometry 
since the effective lattice spacing would diverge due to the existence of spikes \cite{Ambjorn:1997ub}. 
The spike is a ``local'' portion of triangulation in which some space-like edge lengths can be elongated without any end even if fixing the total area of the triangulation (see Fig.~\ref{fig:spike}). 
As shown in Ref. \cite{Ambjorn:1997ub}, 
the average edge length diverges for a finite $n$: 
\[
\left\langle \ell^n \right\rangle_A
= \infty\ ,
\label{eq:spike}
\]
which has been confirmed for various local measures. 
Because of the divergence (\ref{eq:spike}), one cannot define a genuine scale set by the cosmological constant or the total area \cite{Ambjorn:1997ub}. 

As we will see in the next section, in the $2$d Lorentzian QRC the spiky configuration does not exist, 
but another length-divergent configuration called the pinched geometry does exist.

\subsection{Pinched geometries in 2d Lorentzian QRC}
\label{sec:PinchedGeometriesIn2dLorentzianQRC}

\begin{figure}[t]
\centering
\includegraphics[width=3.5in]{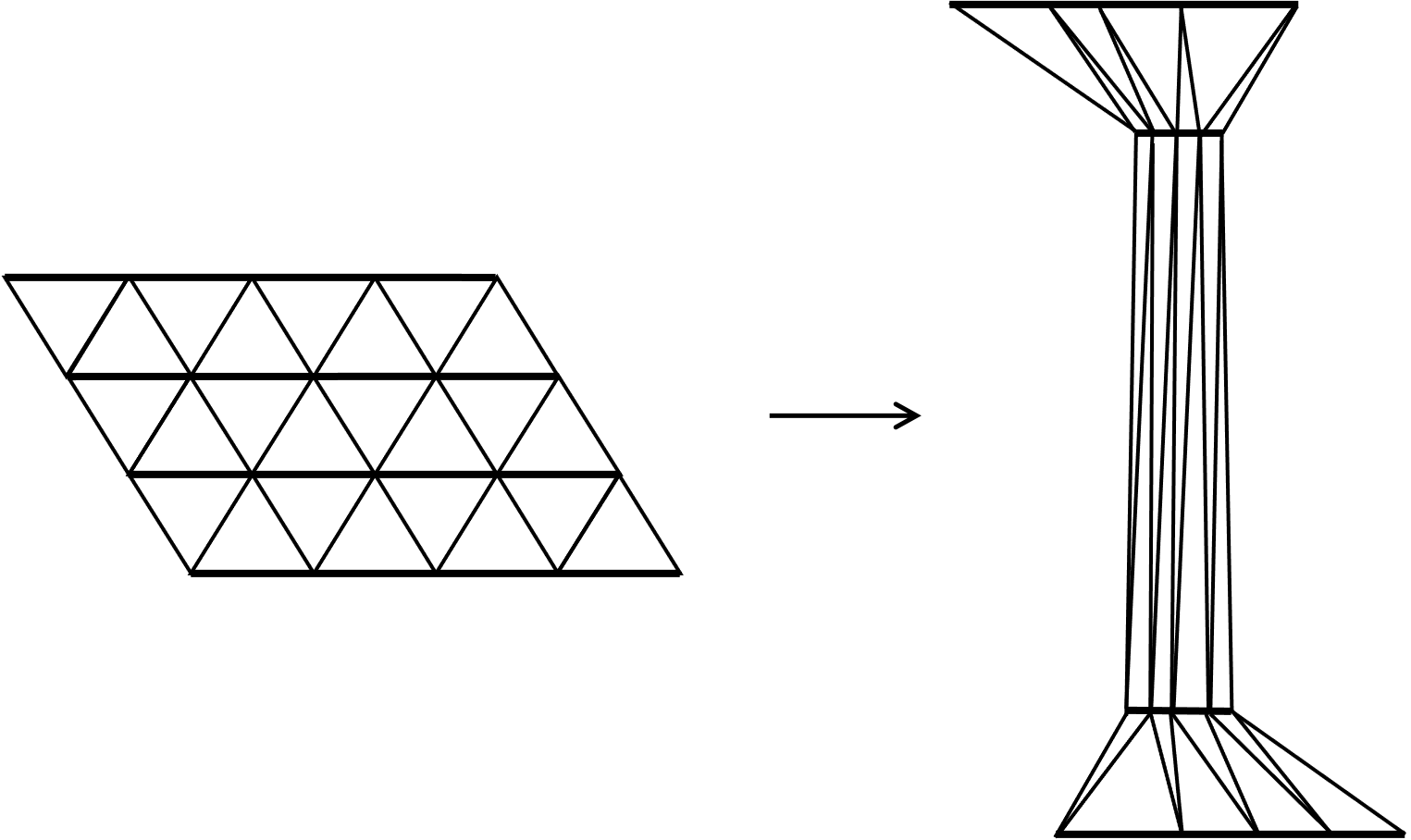}
\caption{The portion of triangulation that can be potentially ``pinched'' with small space-like edge lengths and long time-like edge lengths. The area is unchanged from the left to the right.}
\label{fig:pinch}
\end{figure}

As shown in \cite{Tate:2011ct, Jia:2021deb}, there exists no spike in the $2$d Lorentzian QRC, 
meaning that for any space-like edge length $\sigma$, and for any finite $n$, 
\[
\left\langle \sigma^n \right\rangle_A
< \infty\ . 
\label{eq:nospike}
\]
However, even if fixing the total area, the proper times (time-like edge lengths) can be extended without any end, 
which yields the pinched geometry (see Fig.~\ref{fig:pinch}).  
It would be possible that such pinched geometries yield the divergent effective time-like edge length, i.e. 
\[
\left\langle \tau^n \right\rangle_A
= \infty\ ,
\label{eq:divergenttau}
\]
where $\tau$ is an arbitrary time-like edge length and $n$ a finite number. 

Since it would be possible that the pinched geometry may be entropically suppressed at large $N$, 
in the next section we will construct a tensor-network representation for the $2$d Lorentzian QRC to check if eq.~(\ref{eq:divergenttau}) 
holds or not numerically using the tensor renormalization group (TRG). 
This is the first attempt to apply TRG to QRC, and we bear future applications to other QRC models in mind.

Although the obvious diffeomorphism do not exist in Regge calculus, the emergence of approximate diffeomorphisms has been discussed in Refs.\cite{Hartle:1985wr, Piran:1985is, Hartle:2022ykc}\footnote{
Ref.\cite{Rocek:1982tj} and Ref.\cite{Lehto:1985sb} seem to have different viewpoints other than Hartle's \cite{Hartle:1985wr}.
}. 
In any case, the effective lattice spacing should stay finite in QRC to obtain smooth geometries at large $N$.
This scenario can fail if the spikes exist or if the pinched geometries become dominant. 
As shown in \cite{Tate:2011ct, Jia:2021deb}, the spiky configurations do not appear in $2$d Lorentzian models of QRC. 
This will be confirmed based on our TRG simulations. 
Concerning the pinched geometries, we will show that although there exist configurations of pinched geometry,  
they may be entropically suppressed at large $N$.

\section{Tensor-network representation of the Lorentzian QRC}
\label{sec:TNR}

\subsection{The Lorentzian model with an IR regulator}

The partition function \eqref{eq:partitionfun} defines a Lorentzian model of 2d QRC. 
It is however not well-defined for two reasons. 
Firstly, the Regge action \eqref{eq:Reggeaction} has flat directions that lead to the divergence of the partition function.
The Lorentzian area (\ref{eq:area}) actually stays at a fixed value $A$ 
for $\tau \rightarrow \infty$ where $\tau \vcentcolon = \tau_1=\tau_2$ and $\sigma^2=2\sqrt{\tau^4+A^2}-2\tau^2$.
As shown in Appendix \ref{sec:exact_results}, 
the partition function actually diverges for $N=2$.
In order to lift such flat directions, we introduce an IR regulator $\mu>0$ in the local area $A$ as 
\[
A(\tau_1^2,\tau_2^2,\sigma^2) \to A^{(\mu)} (\tau_1^2,\tau_2^2,\sigma^2)
\vcentcolon =  \frac{1}{4} 
\sqrt{ \sigma^4 + (1+\mu) (\tau^4_1 + \tau^4_2) + 2 ( \sigma^2\tau^2_1 + \sigma^2 \tau^2_2 - \tau^2_1 \tau^2_2 ) }\  , 
\label{eq:mu}
\] 
and take $\mu \to 0$ in the end of the calculation.  
Secondly, for $\beta \le -3/2$, the partition function is ill-defined after introducing $\mu$.  
The integrals in eq.~(\ref{eq:partitionfun}) converge if $\beta > -3/2$ and if $\lambda$ has a positive imaginary part. 
Therefore, in the following we assume 
\begin{align}
\beta > -\frac{3}{2}\ , 
\label{eq:betarange}
\end{align}
and shift the real cosmological constant as $\lambda \to \lambda + i \epsilon$
where $\epsilon$ is a small positive parameter that is taken to zero in the end. 
The $i\epsilon$ prescription allows us to perform an analytic continuation of 
  $Z$ from $i\lambda$ to $-|\lambda|$ as
shown in Appendix \ref{sec:AnalyticContinuation}.

The partition function of the Lorentzian QRC is thus given by
\[
Z
= 
\int^{\infty}_{0} 
\prod^{N_e}_{e=1} d\ell^2_e\ \cdot \prod^{N}_{s =1} \left[ A^{(\mu)}_s  (\{\ell^2\} ) \right]^{\beta}\ \cdot
e^{ - \lambda \sum^N_{s=1} A^{(\mu)}_s (\{\ell^2\})}\ ,
\label{eq:partitionfun2}
\]
for  $\lambda>0$.
Note that an overall factor $C\vcentcolon = i^{N(\beta+3/2)}$ which appears after the analytic continuation is dropped in eq.~\eqref{eq:partitionfun2},
and  the limit of $\epsilon \rightarrow 0$ can be taken safely after the analytic continuation. 
Hereafter we will discuss the positive $\lambda$ case. 
We find that eq.~\eqref{eq:partitionfun2} does not coincide with the partition function of the Euclidean QRC. 
In the analytically continued theory, 
the expectation value of a smooth function ${\cal O}(\{\ell^2\})$ is defined by
\[
\left\langle \mathcal{O} \right\rangle 
= \frac{1}{Z} 
\int [d\ell^2] \ \mathcal{O} (\{\ell^2\}) \ 
e^{ - \lambda \sum^N_{s=1} A^{(\mu)}_s (\{\ell^2\})}\ .
\label{eq:ev}
\]
In the numerical computations shown in the next section, 
eqs.~\eqref{eq:partitionfun2} and \eqref{eq:ev} are used 
as the partition function and the expectation values, respectively.  

The $\lambda$ dependence of $Z$ is extracted as an overall factor.  
Through the change of variables, $\ell^2_e \to \ell^2_e/ \lambda$, we can show that
\begin{align}
Z  = \lambda^{-\alpha} Z|_{\lambda=1}\ ,  
\label{eq:Zformlua1} 
\end{align}
where 
\[
\alpha \vcentcolon =  N (\beta + 3/2)\ .
\label{eq:alpha}
\]
Similarly, we have 
\begin{align}
& \langle {\cal O}_m \rangle  = \lambda^{-m} \langle {\cal O}_m \rangle |_{\lambda=1} \ ,  
\label{eq:Om} 
\end{align}
where $\mathcal{O}_m (\ell_e^2)$ satisfies the relation, $\mathcal{O}_m (\gamma \ell_e^2) = \gamma^m \mathcal{O}_m (\ell^2_e)$ 
for $\gamma>0$.

The fixed-area partition function $Z_A$ 
is defined as 
\[
Z_A
=  \int^{\infty}_0 \prod^{N_e}_{e=1} d\ell^2_e \cdot \prod^N_{s=1}  \left[ A^{(\mu)}_s (\{\ell^2\}) \right]^{\beta} 
\cdot \delta \left(A - \sum^{N}_{s=1} A^{(\mu)}_s (\{\ell^2\} )  \right)\ .
\label{eq:fixed_area_partition_function}
\]
It is easily shown that $Z$ is obtained from the Laplace transform of $Z_A$:
\[
Z 
= \int^{\infty}_0 dA\ e^{-\lambda A} Z_A \ .
\label{eq:laplacez}
\]
The fixed-area expectation value is also defined by
\[
\langle {\cal O} \rangle_A  
=
\frac{1}{Z_A}
 \int^{\infty}_0 \prod^{N_e}_{e=1} d\ell^2_e \cdot \prod^N_{s=1}  \left[ A^{(\mu)}_s (\{\ell^2\}) \right]^{\beta} \, {\cal O}(\{\ell^2\})
\cdot \delta \left(A - \sum^{N}_{s=1} A^{(\mu)}_s (\{\ell^2\} )  \right)\ .
\label{eq:fixed_area_ev}
\]
Note that $\langle\cdot \rangle_A$ does not depend on $\lambda$ and 
is invariant under the analytic continuation 
$\lambda \rightarrow i |\lambda|$.

The change of variables $\ell^2_e \to A \ell^2_e$ tells us that
\begin{align}
Z_A 
= A^{\alpha-1} Z_{A=1}\ , 
\label{eq:Zformula2} 
\end{align}
and 
\begin{align}
\langle {\cal O}_m \rangle_A = A^{m} \langle {\cal O}_m \rangle_{A=1}  \ , 
\label{eq:OmA}
\end{align}
where $A>0$.
We also find that the partition function (\ref{eq:partitionfun2}) is, in fact, proportional to the fixed area partition 
function (\ref{eq:fixed_area_partition_function}): 
Plugging eq.~\eqref{eq:Zformula2} into the Laplace transform (\ref{eq:laplacez}), we obtain  
\[
Z 
= \left( \int^{\infty}_0 dA\ e^{-\lambda A} A^{\alpha-1} \right) Z_1 
= c_A (\lambda,\alpha) Z_A \ ,
\label{eq:eq:laplacez2}
\]
where
\[
c_A (\lambda,\alpha) 
\vcentcolon =   \frac{ A\Gamma [\alpha]  }{ (\lambda A)^{\alpha}}\ .
\label{eq:const}
\]
We also have
\[
\left\langle \mathcal{O}_m \right\rangle_A 
= \frac{(\lambda A)^m \Gamma(\alpha)} {\Gamma(\alpha+m)}
\left\langle \mathcal{O}_m \right\rangle 
\ . 
\label{eq:relation}
\]
Thus we find that $\left\langle \mathcal{O}_m \right\rangle_A$ in the 2d Lorentzian QRC can be obtained 
from the expectation value in the analytically continued theory $\left\langle \mathcal{O}_m \right\rangle$.

The average area of a single triangle can be calculated exactly:
\[
\langle A_{s}^{(\mu)} \rangle = \left( \beta + \frac{3}{2} \right) \frac{1}{\lambda}\ , 
\label{eq:averagearea}
\]
from $\langle A_{s}^{(\mu)} \rangle = - \frac{1}{N} \frac{\partial}{\partial \lambda} {\rm log} Z$.
Note that 
$\langle A_{s} \rangle \vcentcolon =  {\rm lim}_{\mu \rightarrow 0} 
\langle A_{s}^{(\mu)} \rangle =  \left( \beta + \frac{3}{2} \right) \frac{1}{\lambda}$
since the RHS does not depend on $\mu$.
This quantity can be a good benchmark when checking the validity of numerical simulations.

\subsection{Tensor network representation}

\begin{figure}[H]
\centering
\includegraphics[height=4.8cm]{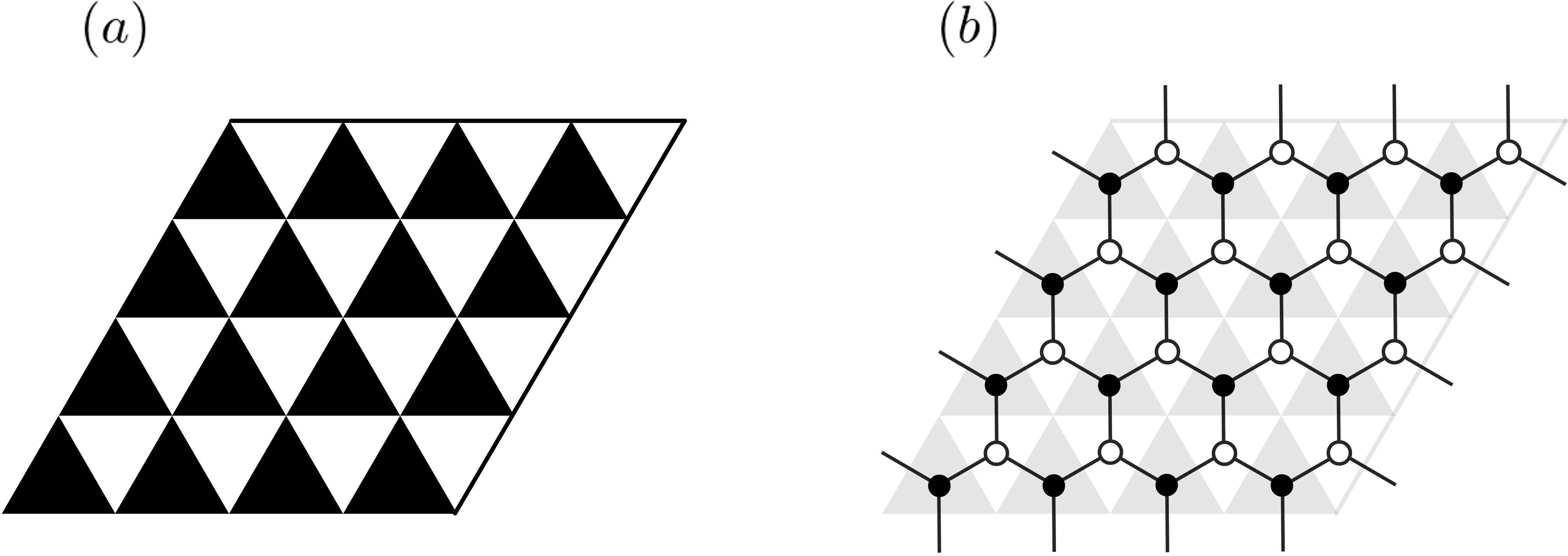}
\caption{Color-coding of a triangulation in black and white $(a)$ and its dual network $(b)$.}
\label{fig:dual}
\end{figure}

\begin{figure}[H]
\centering
\includegraphics[height=5cm]{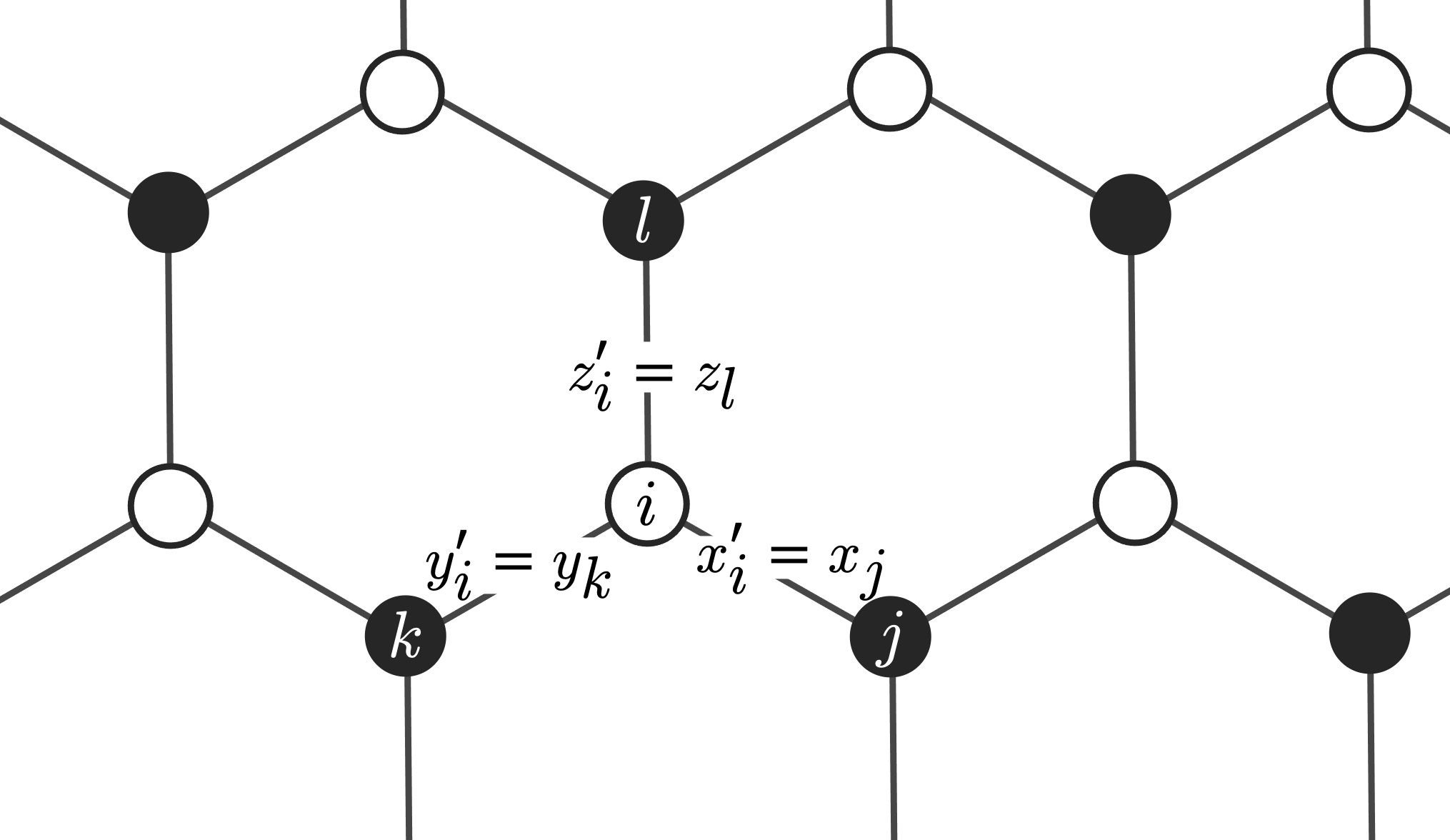}
\caption{Correspondence among indices. 
The honeycomb lattice is the dual network of the black-and-white triangulation (See Fig.~\ref{fig:dual}). 
Link variables $x'_i,y'_i,z'_i$ defined on three links stemmed from the white site $i$ are identified to $x_j, y_k, z_l$ 
emanating from the black sites $j,k,l$, respectively.}
\label{fig:network}
\end{figure}

A tensor network representation of the Lorentzian QRC can be derived 
by noticing 
that the local area is a tensor with three indices $\sigma,\tau_1,\tau_2$. 
Let us identify the upward and downward triangles as the black and white sites, respectively, as shown in Fig.~\ref{fig:dual}.  
Note that the number of black sites (white sites) is $N/2$. 
For simplicity of notation we use $x=\tau_1^2, y=\tau_2^2, z=\sigma^2$. 
Then the partition function is given by 
\[
& Z
=
\int^{\infty}_{0} 
\prod_{i \in b} dx_i dy_i dz_i \ \cdot \prod_{i \in b } \left[ A (x_i,y_i,z_i) \right]^{\beta}\  e^{ - \lambda A (x_i,y_i,z_i) } \nonumber
 \\
& \hspace{3cm} \times \prod_{i \in w} \left[ A (x^\prime_i,y^\prime_i,z^\prime_i) \right]^{\beta} \ e^{ - \lambda A (x^\prime_i,y^\prime_i,z^\prime_i) }\ , 
\]
where 
\[
A(x,y,z) = \frac{1}{4} 
\sqrt{ z^2 + (1+\mu) (x^2 + y^2) + 2 (x+y)z -2xy  }\ .
\label{area_def}
\]
Here $b$ and $w$ are the sets of black and white sites, respectively. 
${x^\prime_j, y^\prime_j, z^\prime_j}$ are 
identified as $x_i, y_i, z_i$ following the correspondence shown in figure \ref{fig:network}. 
We then assign the following tensor to a black site: 
\[
S_{x y z} 
= \left[ A (x,y,z) \right]^{\beta}  e^{-\lambda A(x,y,z)}\ , 
\label{eq:tensor}
\]
while the same tensor with the dashed indices is assign to a white site. 

The partition function is thus expressed as
\[
Z = \int_{0}^{\infty} \prod_{i \in b} dx_i dy_i dz_i \cdot
\prod_{i \in b, j \in w} S_{x_i y_i z_i}  S_{x^\prime_j  y^\prime_j z^\prime_j} \ .
\label{eq:TNR}
\]
Since the contractions are performed among only two tensors, 
the partition function is expressed as a tensor network where the tensors are properly assigned to vertices.  \vspace{1mm}


This tensor network has the infinite bond dimension and a numerical approach is difficult to be applied. 
The situation is similar to the TRG approach to theories with scalar fields 
\cite{Kadoh:2018hqq, Kadoh:2018tis, Kadoh:2019ube}.
In these cases, the Gaussian quadrature is used to discretize the integral of each variable. 
The critical coupling constant estimated in  \cite{Kadoh:2018tis} was consistent with those obtained by other methods, and the Silver Blaze phenomenon was clearly observed in \cite{Kadoh:2019ube}.
So the TRG with a discretization of continuous variables by the Gaussian quadrature properly works.  
In order to use the TRG method to the Lorentzian QRC, we also approximate the integrals in eq.~(\ref{eq:TNR}) by the Gauss-Laguerre quadrature.   

It is known that the Gauss-Laguerre quadrature approximates well the value of integrals of function that damps exponentially, e.g. $e^{-x} f(x)$: 
\[
\int^{\infty}_{0}dx\ e^{-x} f(x) \approx \sum_{x \in S_K} w_K(x) f(x)\ , 
\label{eq:GLQ}
\] 
where $S_K$ is the set of $K$ roots of the Laguerre polynomial $L_K(x)$, 
and $w_{K}(x)$ a weight given by
\[
w_{K}(x) \vcentcolon =  \frac{x}{(K+1)^2 (L_{K+1} (x))^2}\ . 
\label{eq:weight}
\]
Here $K$ is the order that controls the accuracy of approximation.  
For integrals of a general function $h(x)$, we use use the following notation:  
\[
\int^{\infty}_{0} dx\ h(x) 
\approx \sum_{x \in S_K} g_K(x) h(x) \ , 
\label{eq:GLQ2}
\]
where $g_K(x) = w_{K}(x) e^{x}$. 
Note that the efficiency of the approximation (\ref{eq:GLQ2}) depends on a specific form of $h(x)$.

Since the space-like index $z_i$ and the time-like indices $x_i, y_i$ are essentially different indices for the Lorentzian case,  
we take different orders $K_s$ and $K_t$ for the $z$-integrals and the $x,y$-integrals, respectively:
\[
\int^{\infty}_{0} 
\prod_{i \in b} dx_i dy_i dz_i 
\approx 
\prod_{i \in b} \sum_{x_i \in S_{K_t}} \sum_{y_i \in S_{K_t}} \sum_{z_i \in S_{K_s}} 
g_{K_t}(x_i) g_{K_t}(y_i) g_{K_s}(z_i)\ . 
\label{eq:approximation} 
\]
Defining  a rank-three tensor as
\[
T_{xyz} =
\sqrt{g_{K_t}(x) g_{K_t}(y) g_{K_s}(z)} S_{xyz} \ , 
\label{eq:T}
\]
we obtain
\[
Z &\approx  
 {\rm Tr} \left( \prod_{i \in b, j \in w} T_{x_i y_i z_i}  T_{x^\prime_j  y^\prime_j z^\prime_j}     \right)\ , 
\label{eq:GLQ3}
\]
where ${\rm Tr}$ stands for the whole contraction with respect to $x,y,z$ indices.  
Here  the primed indices ${x^\prime_j, y^\prime_j, z^\prime_j}$ are again properly 
identified as $x_i, y_i, z_i$.  
Note that $T$ is a $K_t\times K_t \times K_s$ tensor 
satisfying $T_{xyz} = T_{yxz}$.

\begin{figure}[h]
\centering
\includegraphics[width=1.3in]{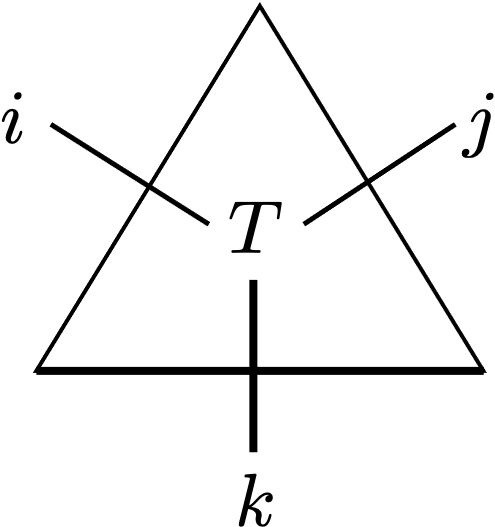}
\caption{A rank-three tensor assigned to a vertex of the dual graph.}
\label{fig:T}
\end{figure}

Eq.~(\ref{eq:GLQ3}) is a tensor-network representation for the $2$d Lorentzian QRC with a finite dimensional tensor $T$, 
i.e. one can introduce a tensor network as the dual graph of the homogeneous triangulation,  
in which the rank-three tensors are assigned to each vertex (see Fig. \ref{fig:T}).  

\begin{figure}[h]
\centering
\includegraphics[width=3.0in]{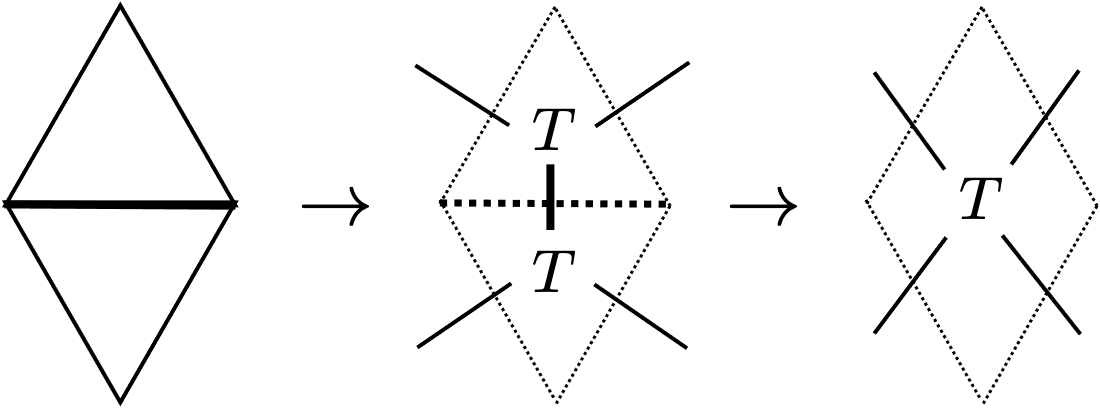}
\caption{How to construct the rank-four tensors: The leftmost, middle and rightmost figures respectively denote
portions of the triangulation, the dual graph consisting of the rank-three tensor and the network made out of the rank-four tensor. 
In the last step, the summation over the index associated with the spatial edge is performed.   
}
\label{fig:T4}
\end{figure}

\begin{figure}[h]
\centering
\includegraphics[width=4.5in]{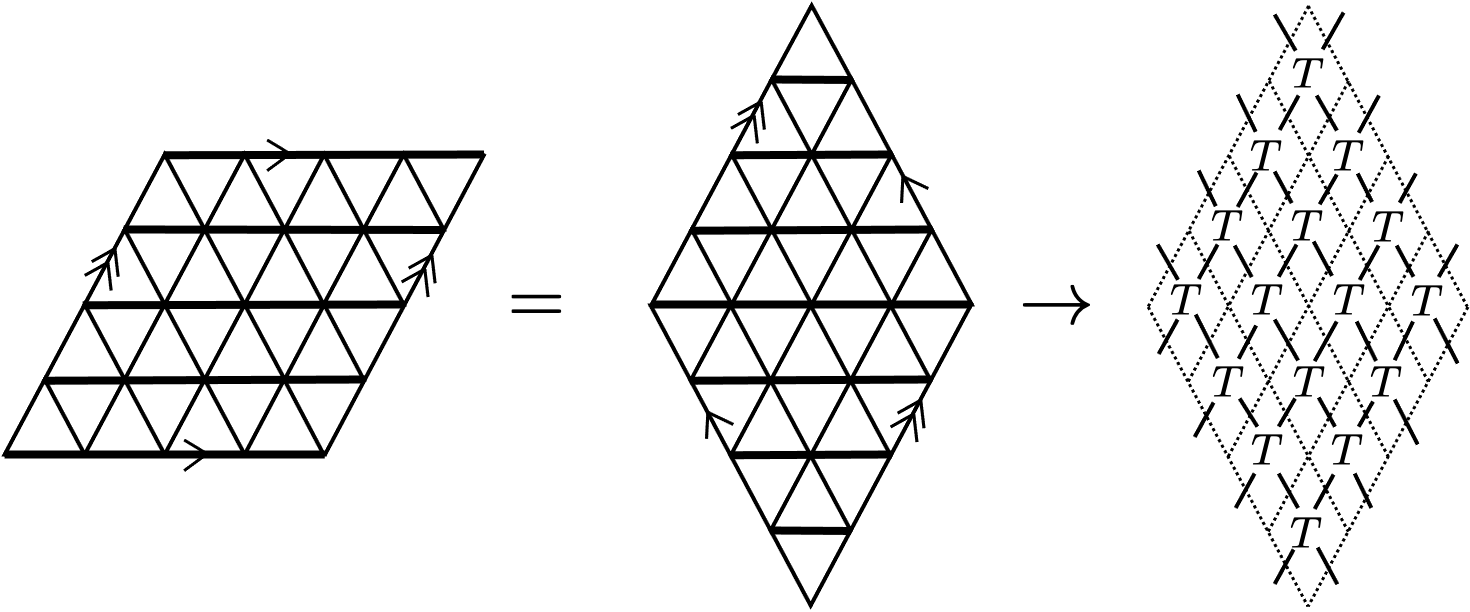}
\caption{How to construct a network made out of the rank-four tensors.}
\label{fig:TRG}
\end{figure}

The tensor network representation \eqref{eq:GLQ3} can be transformed into a more convenient one defined in the square lattice. 
We first perform the summation over the space-like indices $z$'s to  
construct a rank-four tensor. 
The rank-four tensor is given by 
\[
T_{ijkl} \vcentcolon =  
\sum_{z \in S_{K_s}}
T_{ikz} 
T_{jlz} \ .
\label{eq:T4}
\]
For simplicity of notation, we redefine the indices $i,j,k,l$ to be integer running from $1$ to $K_t$.
We thus have
a tensor network of the rank-4 tensor:
\[
Z 
\approx 
 {\rm Tr} \prod_{i \in \Gamma} T_{x_i x^\prime_i y_i  y^\prime_i}\ , 
\label{eq:GLQ4}
\] 
where $\Gamma$ denotes the set of space-like edges.  
Fig.~\ref{fig:T4} and Fig.~\ref{fig:TRG} show how to construct eq.~\eqref{eq:GLQ4} from eq.~\eqref{eq:GLQ3}.

Let $Z$ with insertion of an operator ${\cal O}$  be $I_{{\cal O}}$. 
The expectation value of ${\cal O}$ is then given by a ratio $I_{{\cal O}}/Z$. 
If ${\cal O}$ is local in a sense that it is made only of $\tau_1$, $\tau_2$ and $\sigma$, 
then $I_{{\cal O}}$ can also be expressed as a tensor network with an impurity tensor corresponding to the operator insertion. 
The expectation value is thus evaluated as a ratio of the two tensor networks.

\subsection{Renormalization algorithm}
\label{sec:renormalizationalgorithm}

Once the partition function is represented as a tensor network, 
the numerical value of $Z$ can be evaluated by the tensor renormalization group method. 
In the next section, we use the higher-order TRG (HOTRG) method \cite{Xie:2012} to perform numerical calculations.  
We review the $2$d HOTRG for the readers unfamiliar with the method. 

The renormalization based on the HOTRG is carried out for the $x$ and $y$ directions alternately.
Suppose that the tensor network is given by a form of eq.~\eqref{eq:GLQ4}
with the bond dimension $D$. 
Without loss of generality, we consider the renormalization along the $y$ axis.
We define a tensor $M$ from two tensors: 
\begin{align}
M_{X X^\prime y y^\prime } \vcentcolon = \sum_{d=1}^D T_{x_1x_1^\prime dy^\prime} T_{x_2 x_2^\prime y d}\ , 
\end{align}
where $X = x_1\otimes x_2$ and   $X^\prime = x_1^\prime\otimes x_2^\prime$.
Note that the dimension of $X,X^\prime$  is $D^2$ while that of $y,y^\prime$ is $D$.

Define a matrix $M^\prime_{X, X^\prime yy^\prime} $ as $M^\prime_{X,X^\prime yy^\prime} \vcentcolon = M_{XX^\prime yy^\prime} $
specifying the column and row indices separated by a comma. 
We then diagonalize $K=M^\prime M^{\prime \dag}$ as 
\[
K =U_L \Lambda_L U^\dag_L\ , 
\]
where $\Lambda_L$ is  a diagonal matrix whose diagonal entries are eigenvalues sorted in descending order, and 
$U_L$ are corresponding eigenvectors. Similarly, using a different matrix representation of $M$ as 
$M^\prime_{X^\prime, X yy^\prime} \vcentcolon = M_{XX^\prime yy^\prime}$, 
we can obtain different eigenvalues $\Lambda_R$ and eigenvectors $U_R$, which is done by 
diagonalizing $K=M^\prime M^{\prime \dag}$ as $K =U_R \Lambda_R U^\dag_R$.
We choose an isometry $U$ used in the renormalization by
comparing the residuals:
\begin{align}
U=\left\{
\begin{array}{cc}
U_L   &   \ {\rm for} \ \  \epsilon_L< \epsilon_R  \\
U_R^\ast  & \ {\rm for} \ \ \epsilon_L \ge \epsilon_R
\end{array}
\right. \ , 
\end{align}
where $\epsilon_L = \sum_{i>D} (\Lambda_L)_{ii}$ and $\epsilon_R = \sum_{i>D} (\Lambda_R)_{ii}$. 

\begin{figure}[H]
\centering
\includegraphics[width=5in]{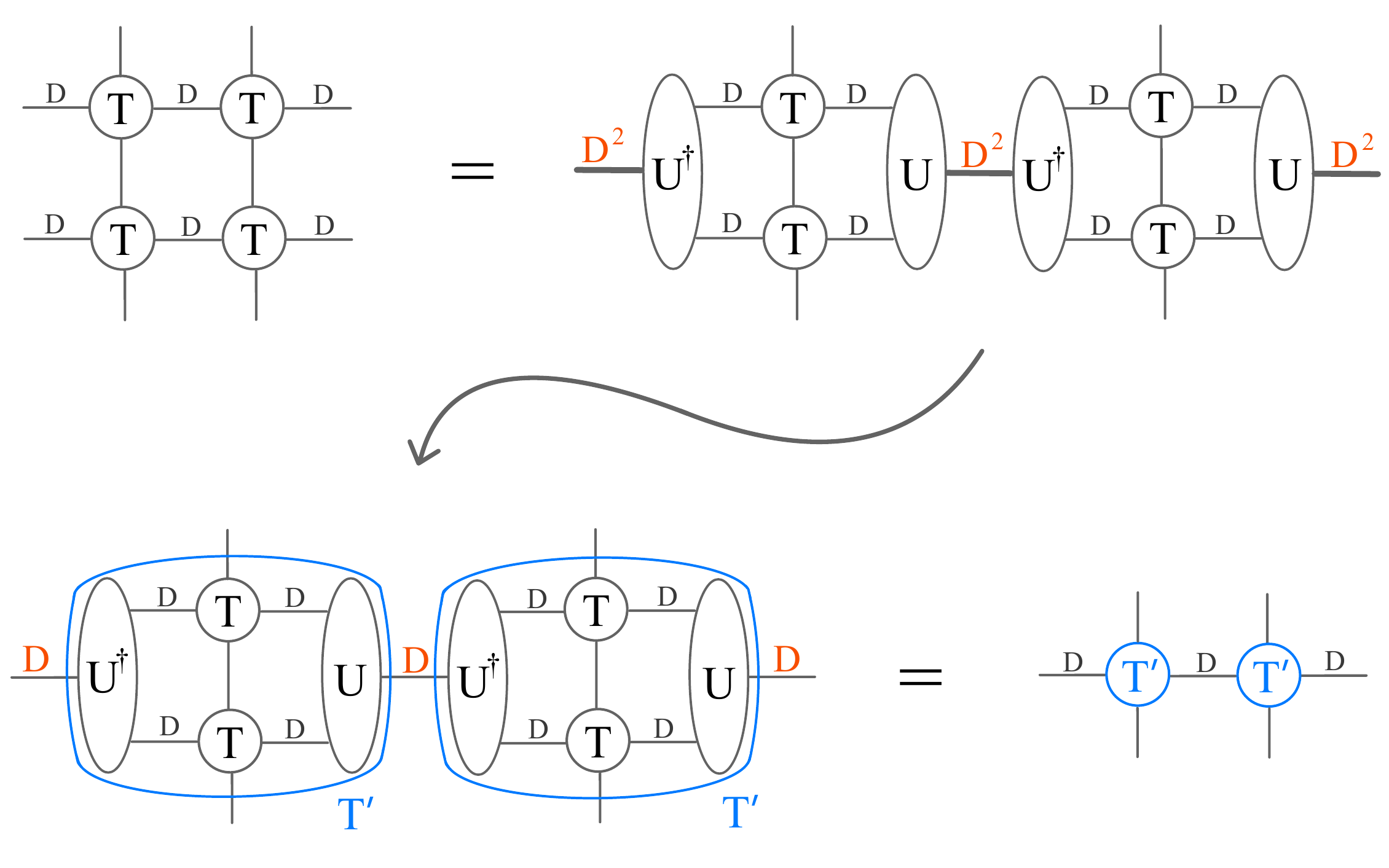}
\caption{
Higher order TRG: 
The top-left figure is a tensor network before the renormalization. 
Circled $T$'s denote the rank-four tensors and the links carry indices running from $1$ to $D$.
In the first equality, the unities, i.e. $U^{\dagger}U=1$ whose indices run from $1$ to $D^2$, 
are inserted. 
From the top-right figure to the bottom-left figure, 
the thick links running from $1$ to $D^2$ between $U$ and $U^{\dagger}$, are deformed  
such that their indices run from $1$ to $D$. 
In the second equality, contractions of the indices of the tensors surrounded by the blue circles are partially performed, resulting in the new tensors $T'$'s. 
The same procedure should be implemented for the vertical links, and this cycle persists until the only one tensor is left. 
This manipulation would give a good approximation if the unitary $U$ is properly chosen. 
}
\label{fig:HOTRG1}
\end{figure}

Using the isometry $U$, two tensors are renormalized into a new tensor $T^\prime$ as 
\begin{align}
T^\prime_{yy^\prime xx^\prime} = \sum_{i,j=1}^{D^2} U^{\dag}_{xi} M_{ij y y^\prime } U_{jx^\prime}.
\label{ren_tensor}
\end{align}
We again reach the tensor network \eqref{eq:GLQ4} of the bond dimension $D$ 
by truncating the bond dimension of $T^\prime$ to be $D$.
The truncation is expected to work effectively as long as the eigenvalues show a clear hierarchy.
Fig.~\ref{fig:HOTRG1} shows one renormalization step.  

The number of tensors is reduced by half after a tensor renormalization. 
For $N=2\times 2^p \times 2^p$, the partition function can be approximated by a single tensor if repeating 
renormalizations $2p$ times. 
At the final step, the numerical value of $Z$ can be evaluated approximately
by performing the remaining two contractions. 

\section{Numerical results}
\label{sec:NumericalResults}

The contribution of the pinched geometry is evaluated using the TRG method in the limit where the  number of triangles is large. 
The partition function of the Lorentzian QRC is expressed as a tensor network \eqref{eq:TNR}. 
Similarly, the numerator of a one-point function 
is expressed as a tensor network with an  impurity tensor. 
The HOTRG method is used to evaluate the numerical values of these tensor networks.  
One point function such as $\langle A_s \rangle$, $\langle l_s^2 \rangle$ and $\langle l_t^2 \rangle$ can be obtained
from the ratio of the two tensor networks.

We can set $\lambda = 1$ without loss of generality because $\lambda$ is factored out 
from the one point functions as shown in eq.~\eqref{eq:Om}.
The expectation values then depend on the six parameters, $N$, $\beta$, $\mu$, $D$, $K_s$ and $K_t$ where  
$N$ is the number of simplices,
$\beta$ the parameter that appears in the measure \eqref{eq:measure},
$\mu$ the IR regulator,
$D$ the bond dimension of the renormalized tensor, 
and 
$K_s$ and $K_t$ are the orders of the Gauss-Laguerre quadrature for the two kinds of index associated with the space-like and the time-like links, respectively.
We select four numerical values for the number of triangles such that $N=2\times 2^p \times 2^p$ with $p=0,1,2,10$,  
which is expressed in the form of $N=2^{2p+1}$.  
The main results are shown in the case of $\beta=0$ and $\beta=1$,  
and 
$\mu$ typically takes values in $10^{-5} \lesssim \mu \lesssim 1$.  
We basically fix the parameters such that $K_s=100$, $K_t=50$ and $D=30$, 
and change them to check the convergence of the results.

\begin{figure}[H]
\centering
\includegraphics[width=8cm]{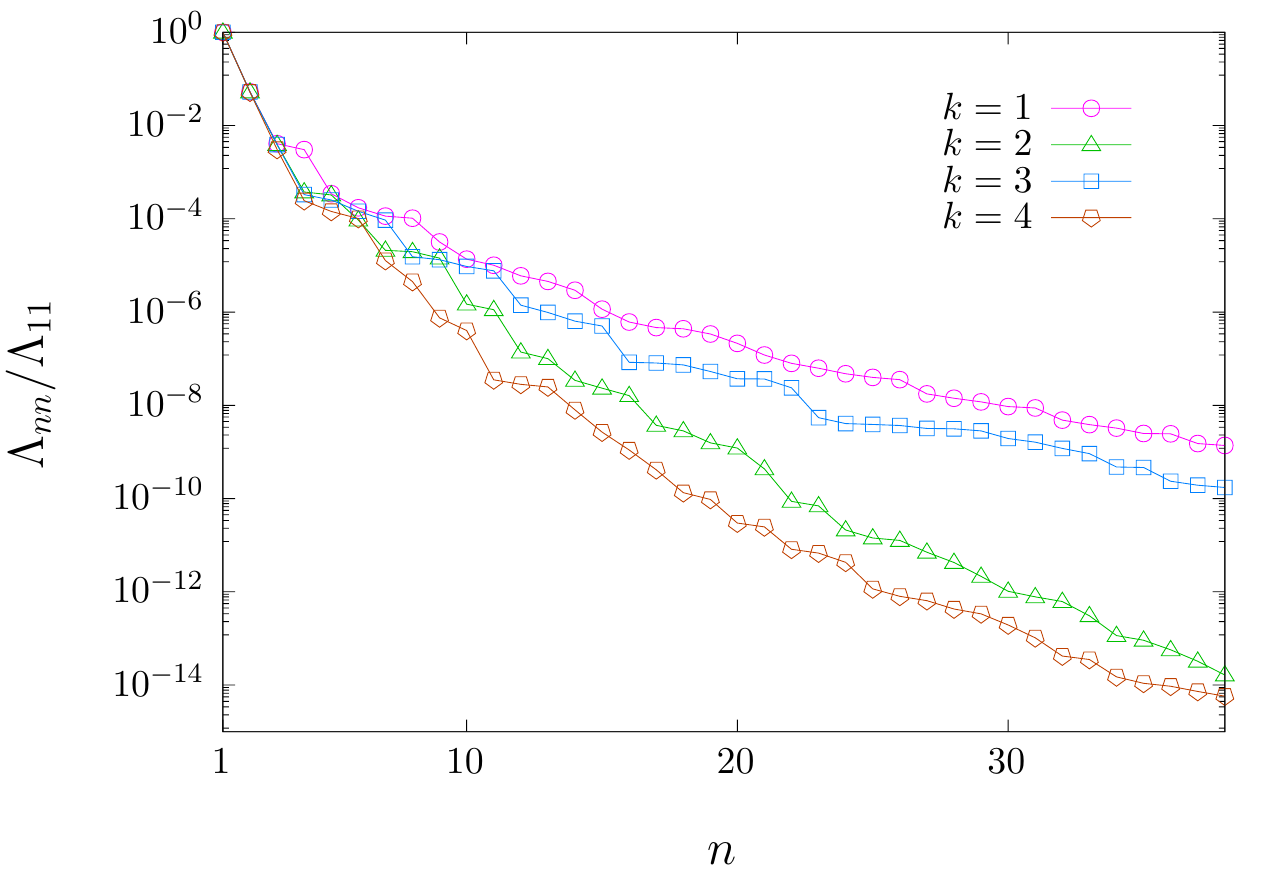}
\caption{The $n$-th eigenvalue normalized by the largest eigenvalue at the $k$-th renormalization where 
$\beta=1$, $\mu=0.3$, $(K_s,K_t)=(100,50)$ and $D=30$. They show clear hierarchies.}
\label{fig:SV}
\end{figure}

In the HOTRG method, the tensor network is renormalized in order that the 
$D$ largest eigenvalues of two tensors are included into a renormalized tensor as
reviewed in Section \ref{sec:renormalizationalgorithm}. 
The method is expected to be effective when the eigenvalues have a clear hierarchy. 
In Fig.~\ref{fig:SV}, obtained eigenvalues are plotted for the first 4 renormalizations.
They show a clear hierarchy which implies that truncated eigenvalues are less important.

\begin{figure}[H]
\centering
\includegraphics[width=8cm]{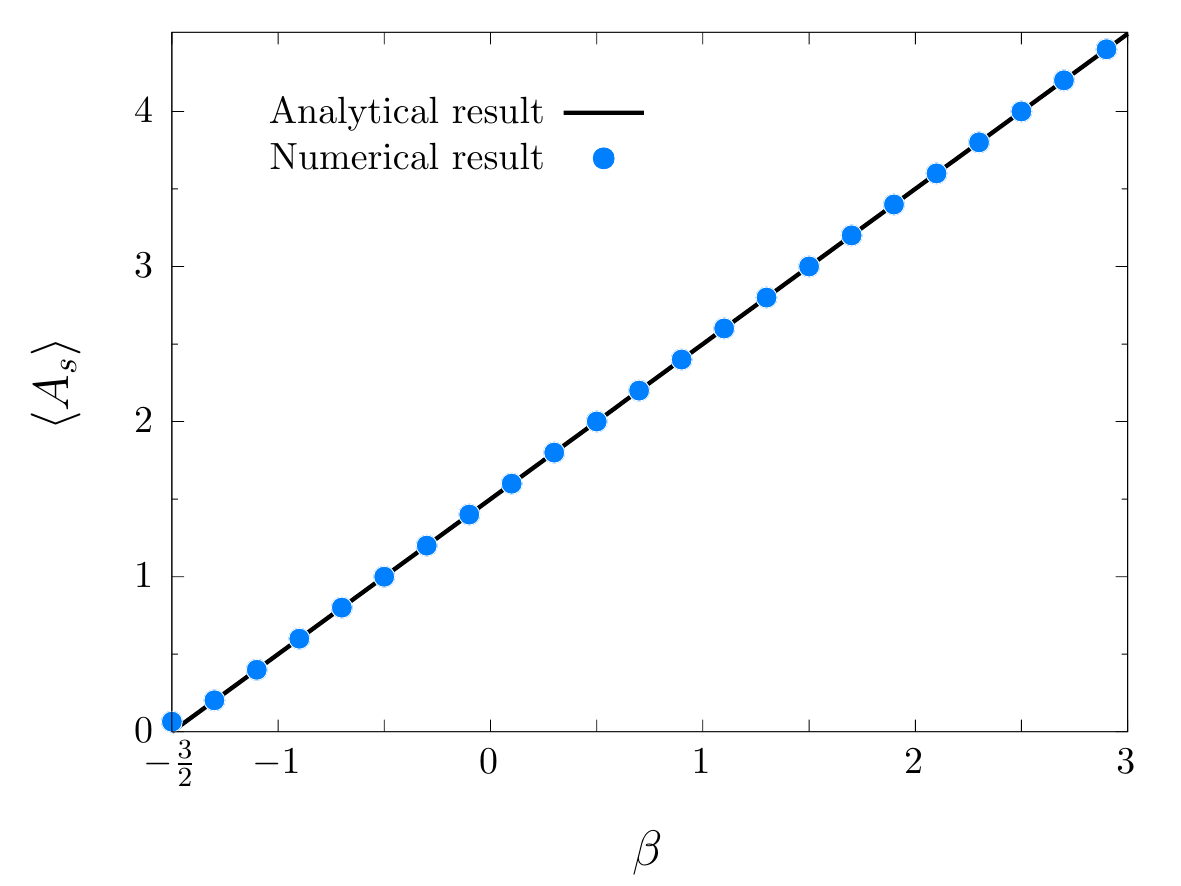}
\caption{The $\beta$ dependence of $\langle A_s \rangle$ at $N=2^{21}$, $\mu=0.3$, $(K_s,K_t)=(100,50)$ and $D=30$. The solid line denotes
the exact values.}
\label{fig:Area}
\end{figure}

\begin{figure}[H]
\centering
\begin{minipage}{0.33\hsize}
 \begin{center}
  \includegraphics[width=5cm]{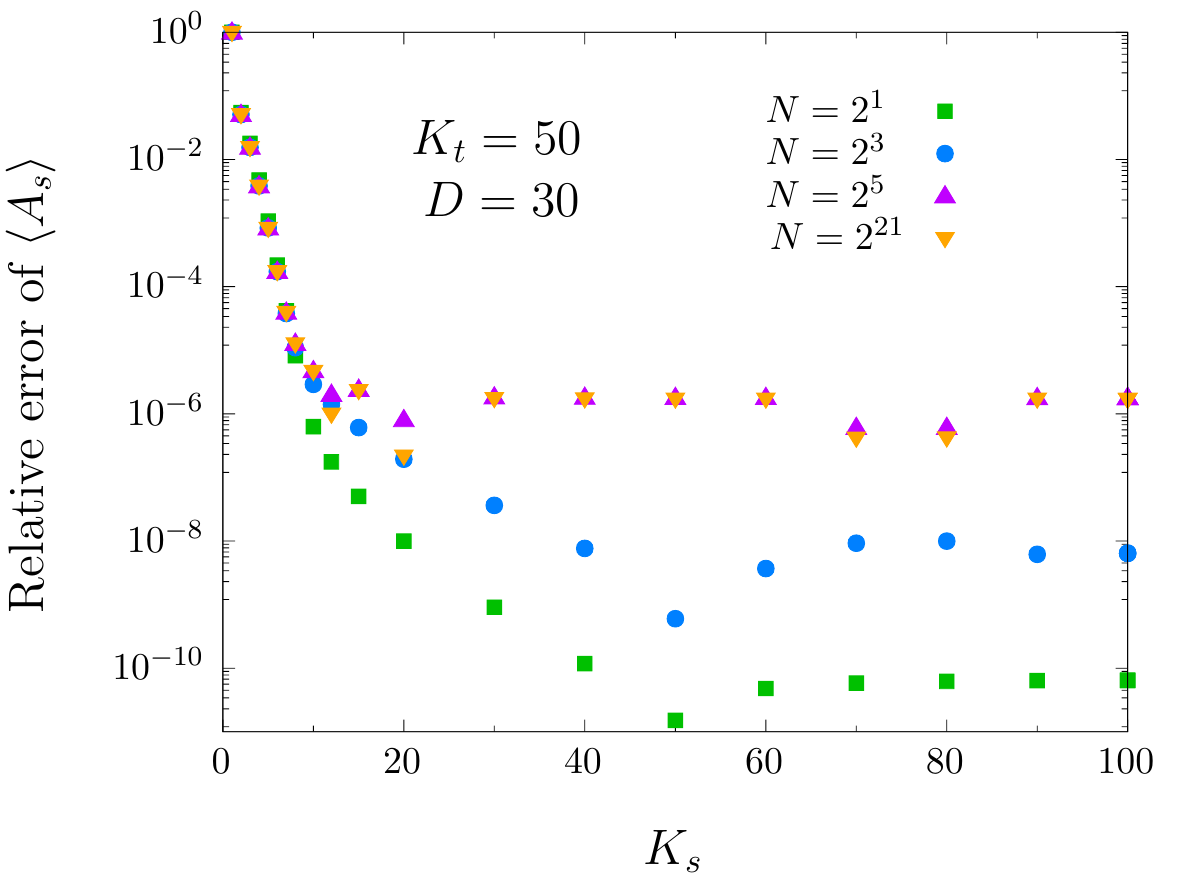}
 \end{center}
\end{minipage}%
\begin{minipage}{0.33\hsize}
 \begin{center}
  \includegraphics[width=5cm]{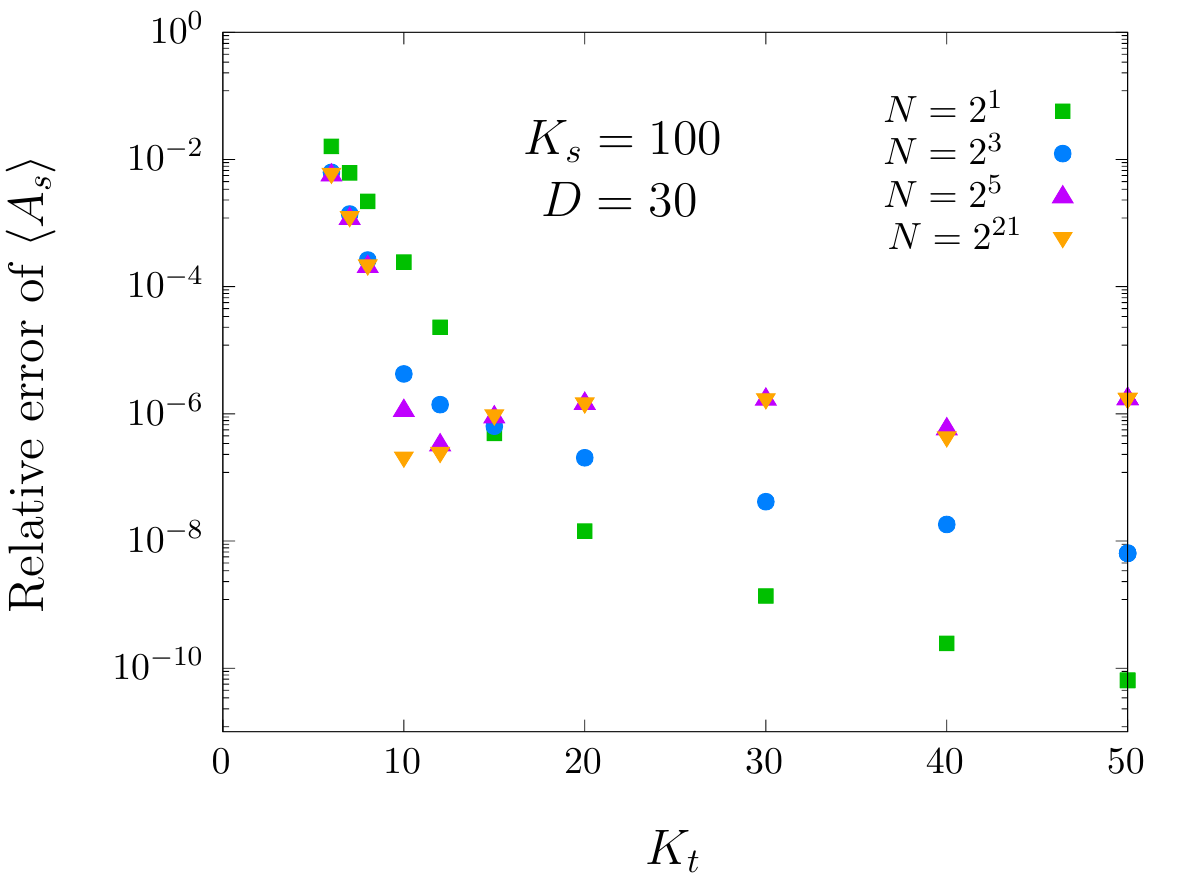}
 \end{center}
\end{minipage}%
\begin{minipage}{0.33\hsize}
 \begin{center}
  \includegraphics[width=5cm]{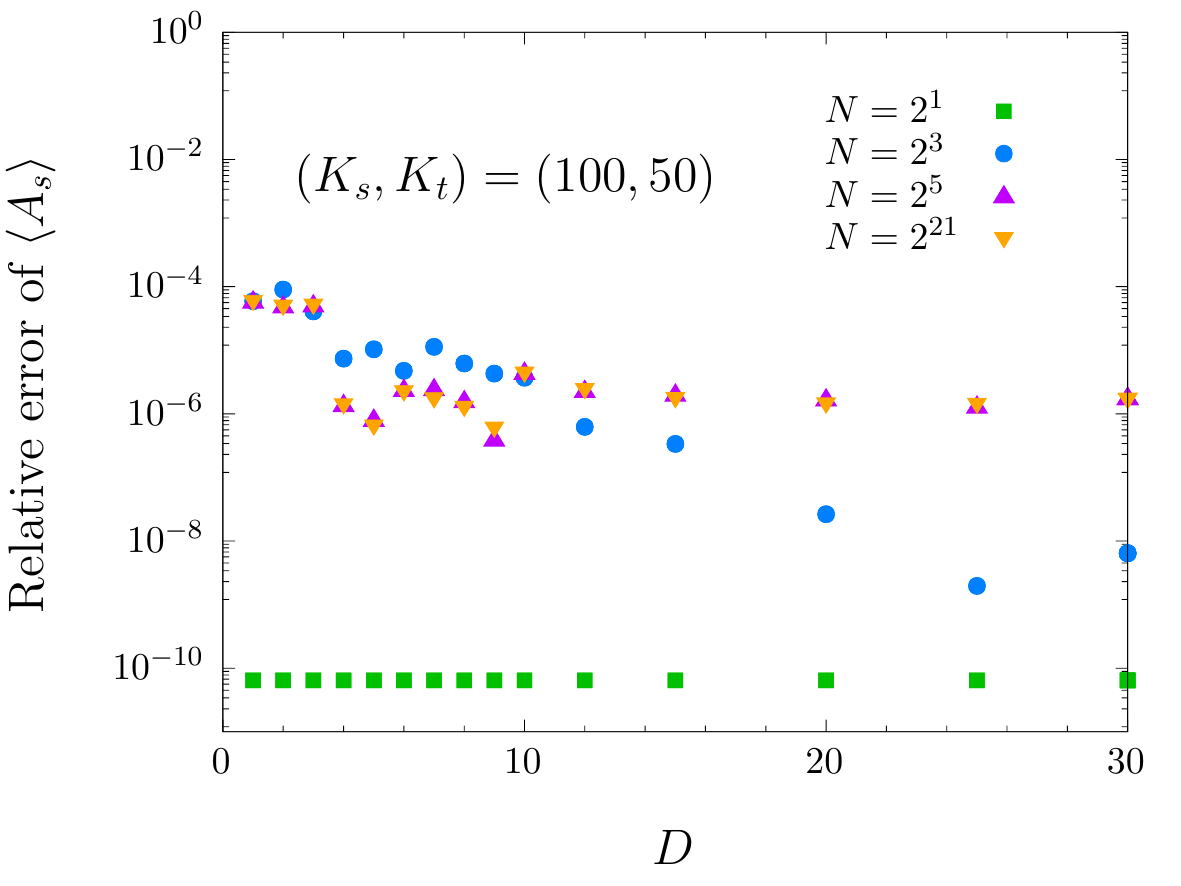}
 \end{center}
\end{minipage}
\caption{$\Delta A$ at $\beta=1$, $N=2,2^3,2^5,2^{21}$ and $\mu=0.3$.
The leftmost figure shows the $K_s$ dependence of $\Delta A$ at $K_t=50$ and $D=30$. 
The middle figure shows the $K_t$ dependence of $\Delta A$ at $K_t=100$ and $D=30$. 
The rightmost figure shows the $D$ dependence of $\Delta A$ at $K_s=100$ and $K_t=50$. 
Increasing $K_s$, $K_t$ and $D$ improves the accuracy. }
\label{fig:error}
\end{figure}

Fig.~\ref{fig:Area}  shows the expectation value of the area $A_s$ given by eq.~\eqref{eq:mu} against $\beta$ 
for $N=2^{21}$, $\mu=0.3$, $K_s=100$, $K_t=50$ and $D=30$. 
As clearly seen in the figure, the numerical results nicely
reproduce the exact value $\langle A_s \rangle= \beta+3/2$, presented in eq.~\eqref{eq:averagearea}. 
Fig.~\ref{fig:error} shows the relative error of $\langle A_s \rangle$ 
defined by $\Delta A = \left|1 - \langle A_s \rangle /(\beta+3/2)\right|$ 
against $K_s,K_t$ and $D$ for $\beta=1$ and $\mu=0.3$. 
The accuracy is improved by increasing $K_s,K_t$ and $D$.
These results show that the TRG calculation works well for the analytically continued 2d Lorentzian QRC.

\begin{figure}[H]
\centering
\includegraphics[width=8cm]{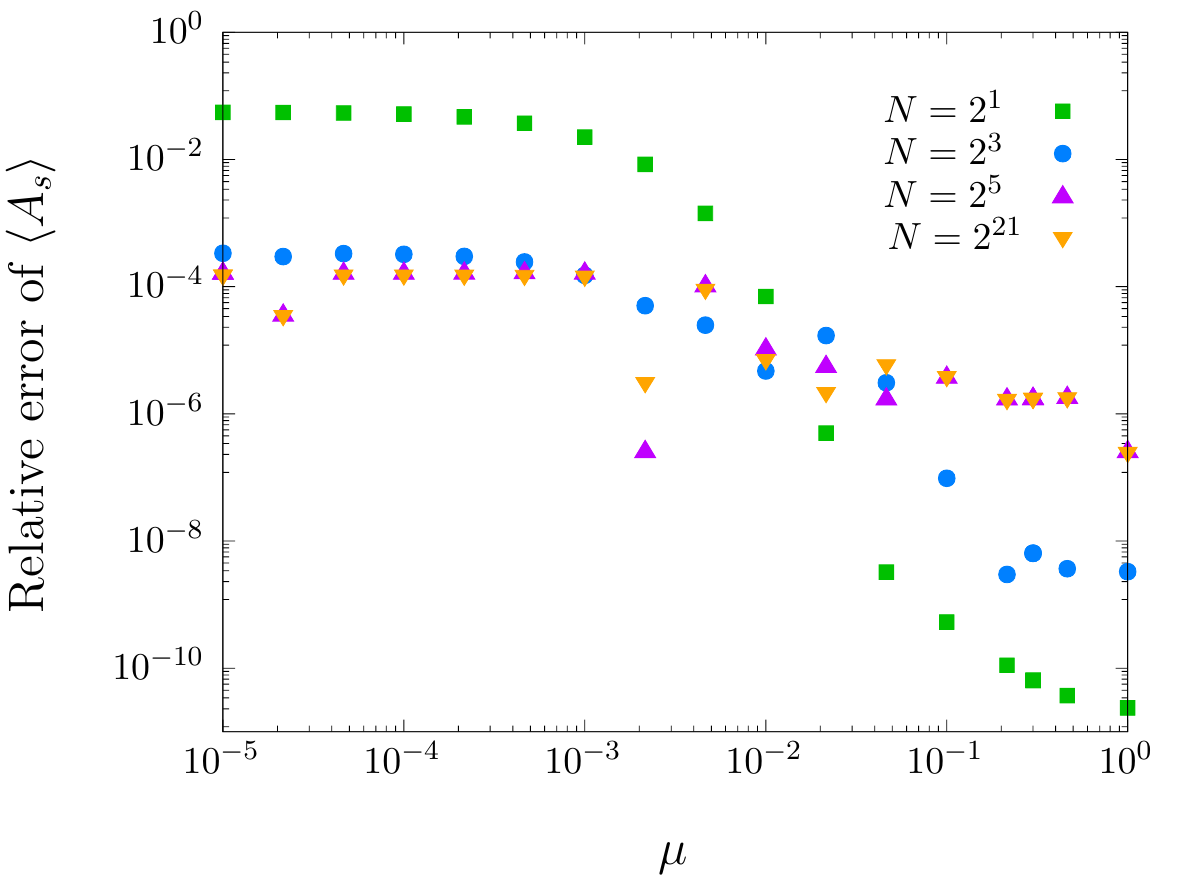}
\caption{The $\mu$ dependence of $\Delta A$ at $\beta=1$, $N=2,2^3,2^5,2^{21}$, $(K_s,K_t)=(100,50)$ and $D=30$. }
\label{fig:error-mu}
\end{figure}

\begin{figure}[H]
\centering
\includegraphics[width=8cm]{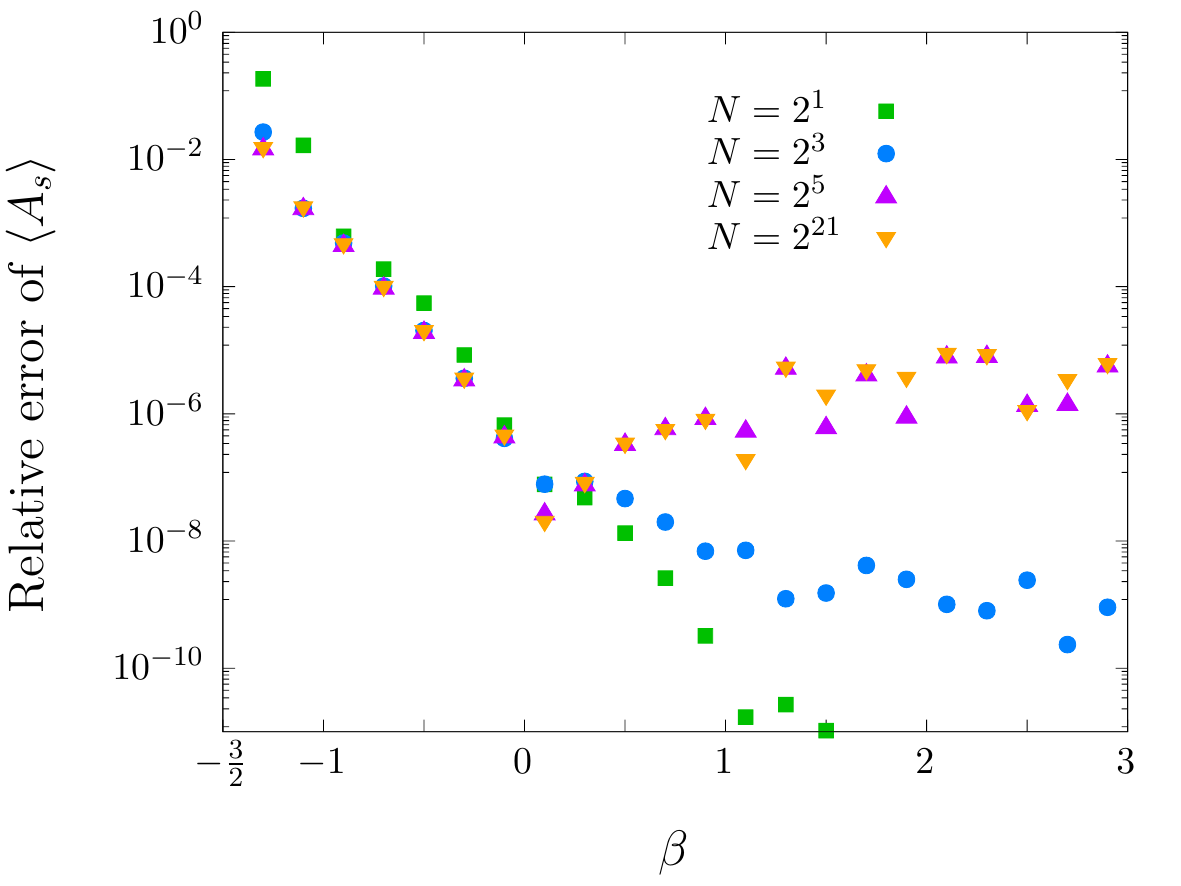}
\caption{The $\beta$ dependence of $\Delta A$ at $\mu=0.3$, $N=2,2^3,2^5,2^{21}$, 
$(K_s,K_t)=(100,50)$ and $D=30$. The accuracy gets worse as $\beta$ approaches $-3/2$. }
\label{fig:error-beta}
\end{figure}

The IR regulator introduced to make the partition function well defined has to be close to zero at the end of the calculation.  
In Fig.~\ref{fig:error-mu}, we find that $\langle A_s \rangle$ are obtained with an error of less than $0.1$ percent 
for  $10^{-5} \lesssim \mu \lesssim 1$ when $N=2^3,2^5,2^{21}$. 
However, when $N=2$,
the error is two order of magnitude larger for $\mu \lesssim 10^{-3}$.
Fig.~\ref{fig:error-beta} shows the $\beta$ dependence of $\Delta A$.
The accuracy decreases as $\beta$ approaches $-3/2$, which corresponds to the scale invariant measure. 
In the following, we present results for $\beta=0$ and $\beta=1$ in the range of
$\mu \gtrsim 10^{-5}$. However, note that the results of $N = 2$ are not converging for $\mu \lesssim 10^{-3}$.

\begin{figure}[H]
\centering
\includegraphics[width=8cm]{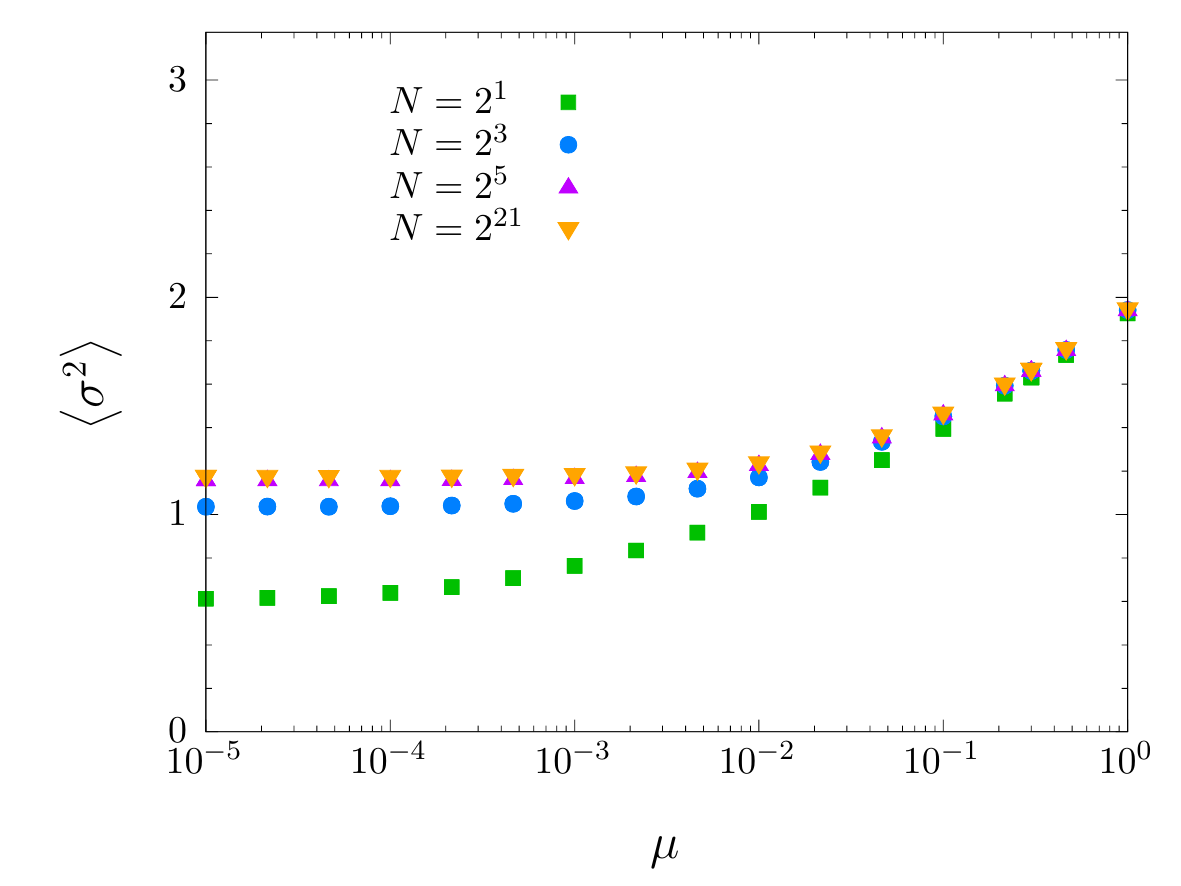}
\includegraphics[width=8cm]{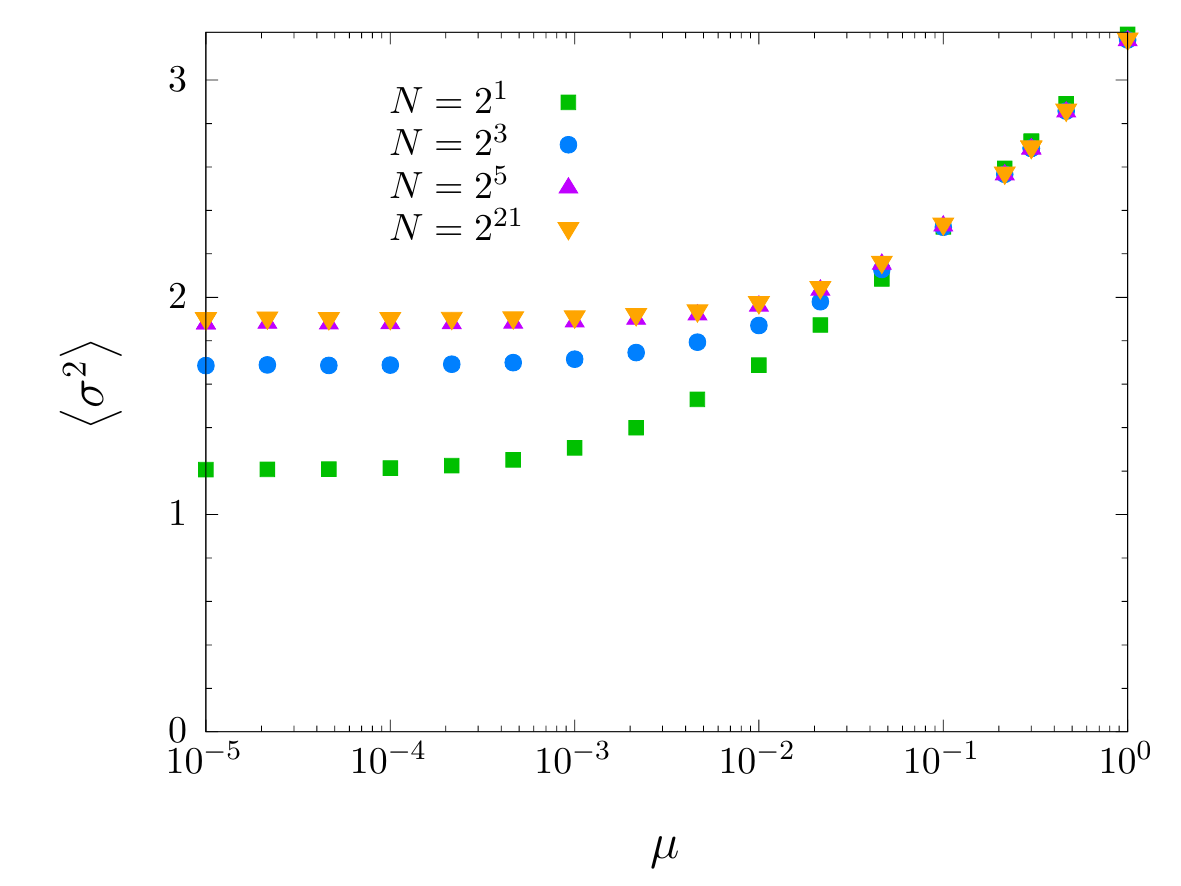}
\hspace{1cm}
\caption{The $\mu$ dependence of $\langle \sigma^2\rangle$ at $\beta=0$ (Left) and $\beta=1$ (Right), $N=2,2^3,2^5, 2^{21}$, $(K_s,K_t)=(100,50)$, and $D=30$. 
}
\label{fig:s-mu}
\end{figure}

\begin{figure}[H]
\centering
\begin{minipage}{0.33\hsize}
  \begin{center}
    \includegraphics[width=5cm]{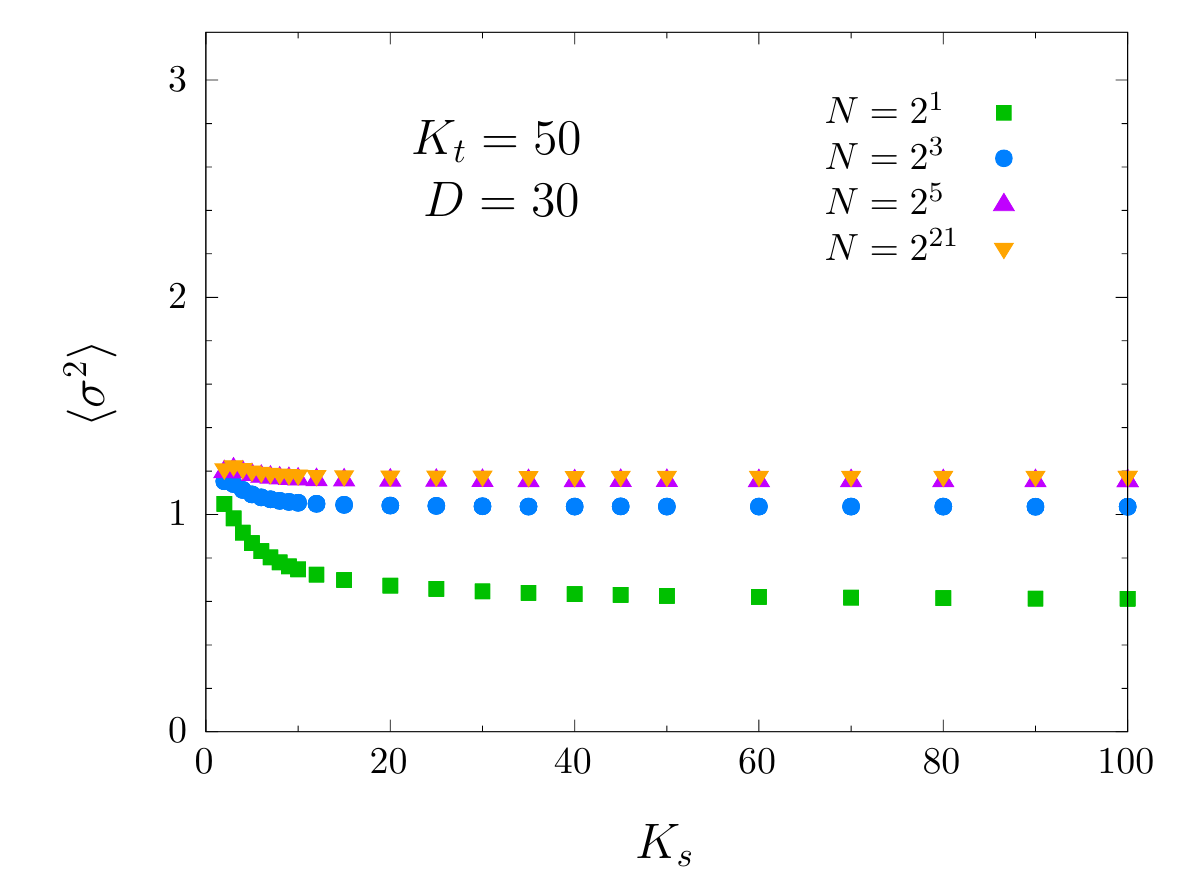}
  \end{center}
\end{minipage}%
\begin{minipage}{0.33\hsize}
  \begin{center}
    \includegraphics[width=5cm]{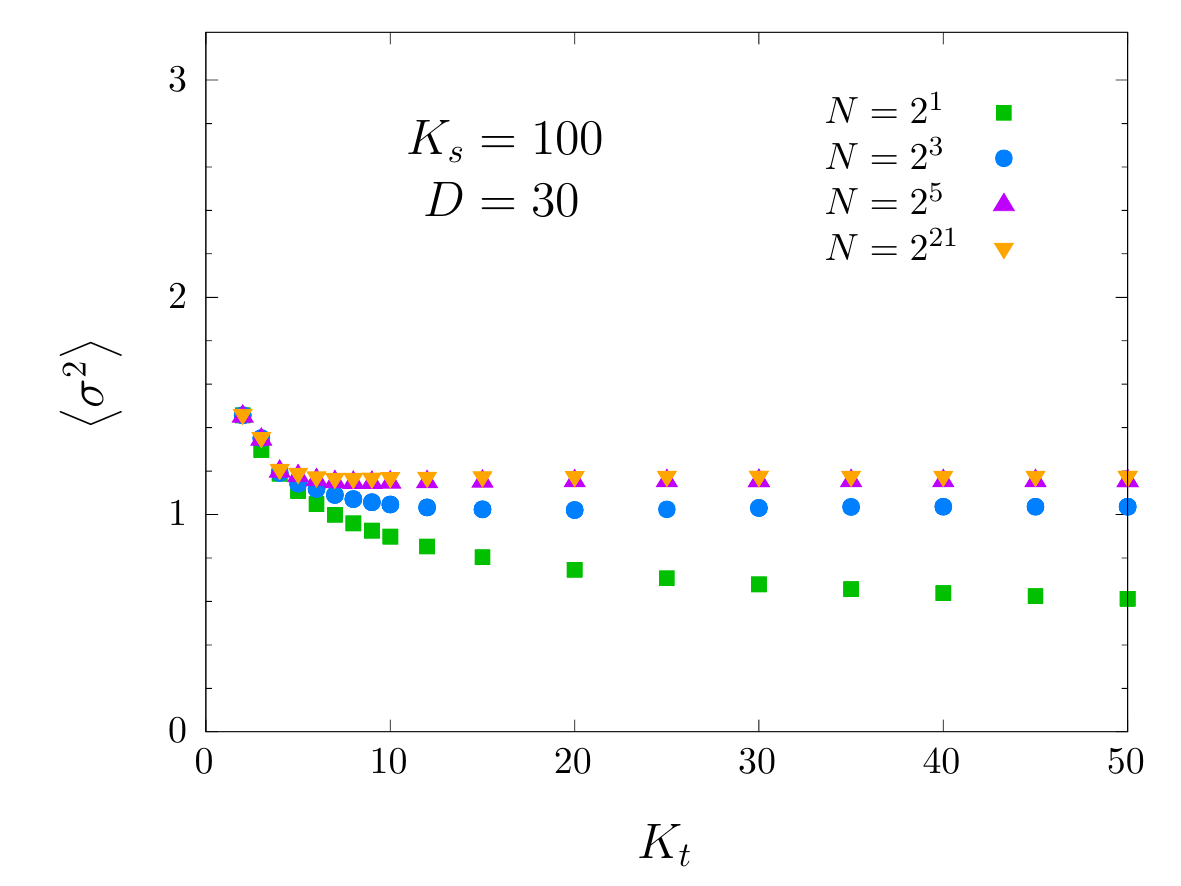}
  \end{center}
\end{minipage}%
\begin{minipage}{0.33\hsize}
  \begin{center}
    \includegraphics[width=5cm]{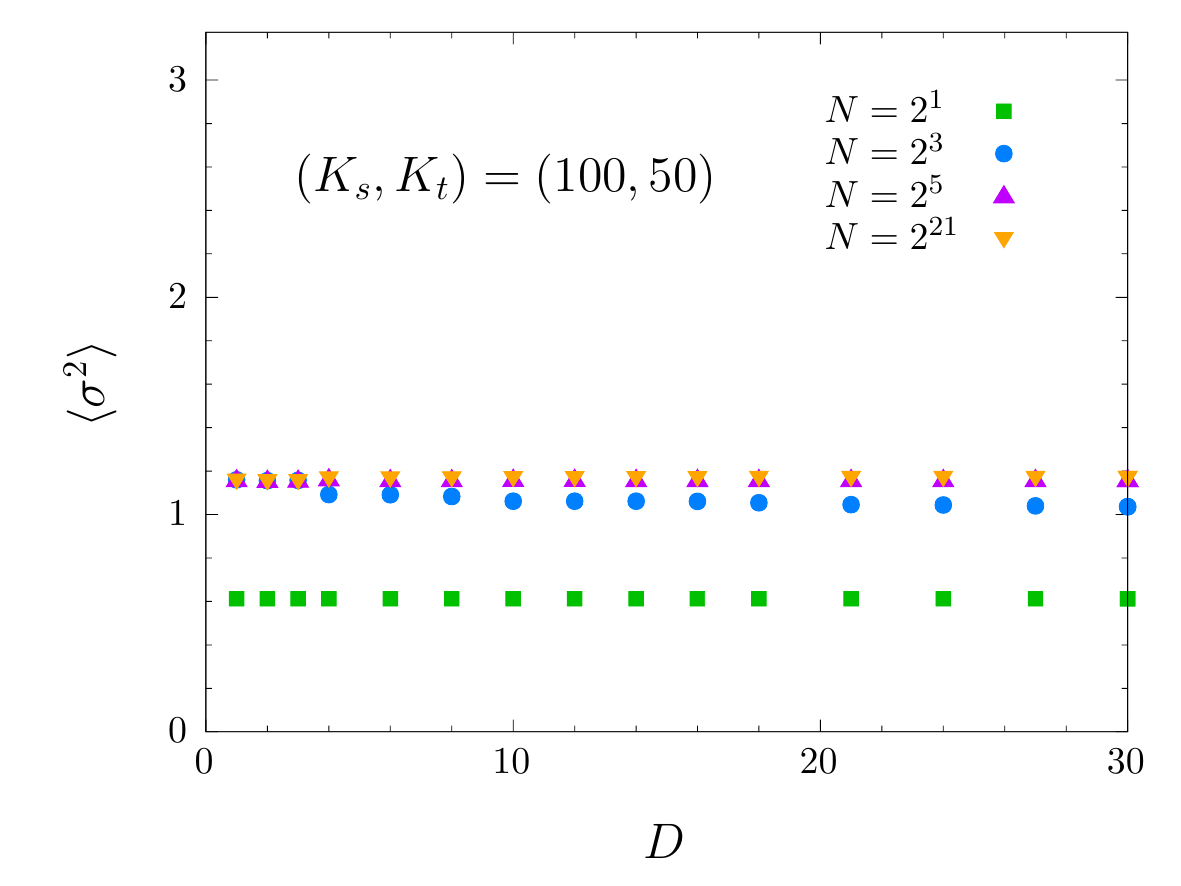}
  \end{center}
\end{minipage}
\caption{The $K_s, K_t$ and $D$ dependence of $\langle \sigma^2 \rangle$ at $\beta=0$, $N=2,2^3,2^5,2^{21}$ and $\mu=10^{-5}$.}
\label{fig:ls_B0}
\end{figure}

\begin{figure}[H]
\centering
\begin{minipage}{0.33\hsize}
  \begin{center}
    \includegraphics[width=5cm]{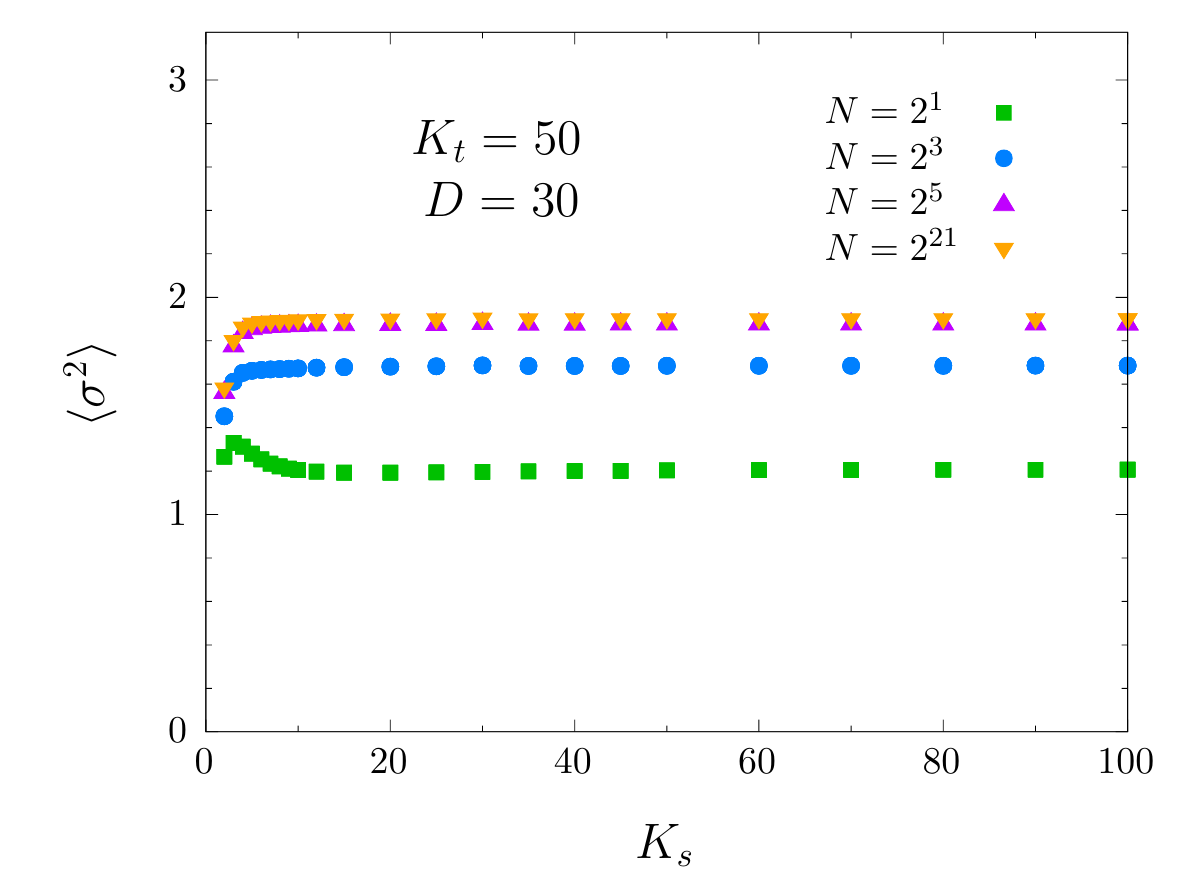}
  \end{center}
\end{minipage}%
\begin{minipage}{0.33\hsize}
  \begin{center}
    \includegraphics[width=5cm]{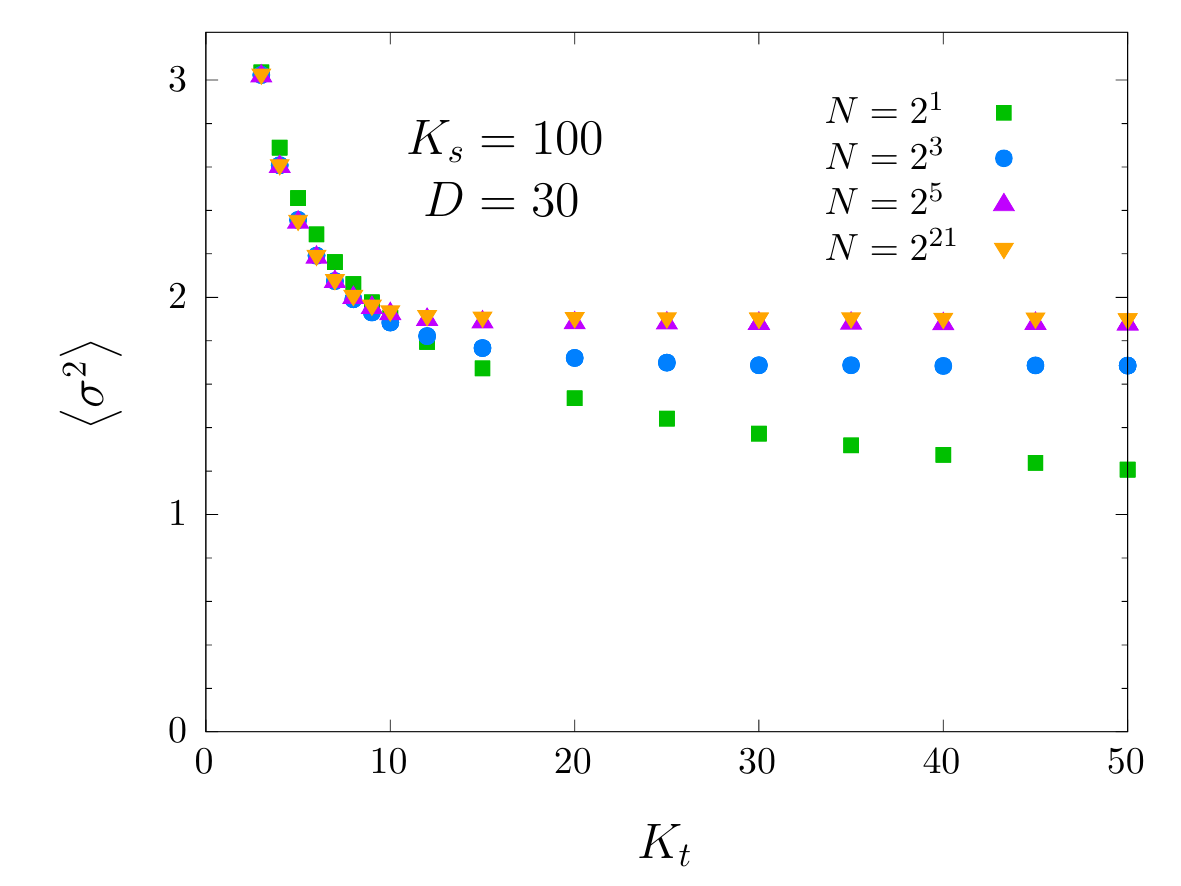}
  \end{center}
\end{minipage}%
\begin{minipage}{0.33\hsize}
  \begin{center}
    \includegraphics[width=5cm]{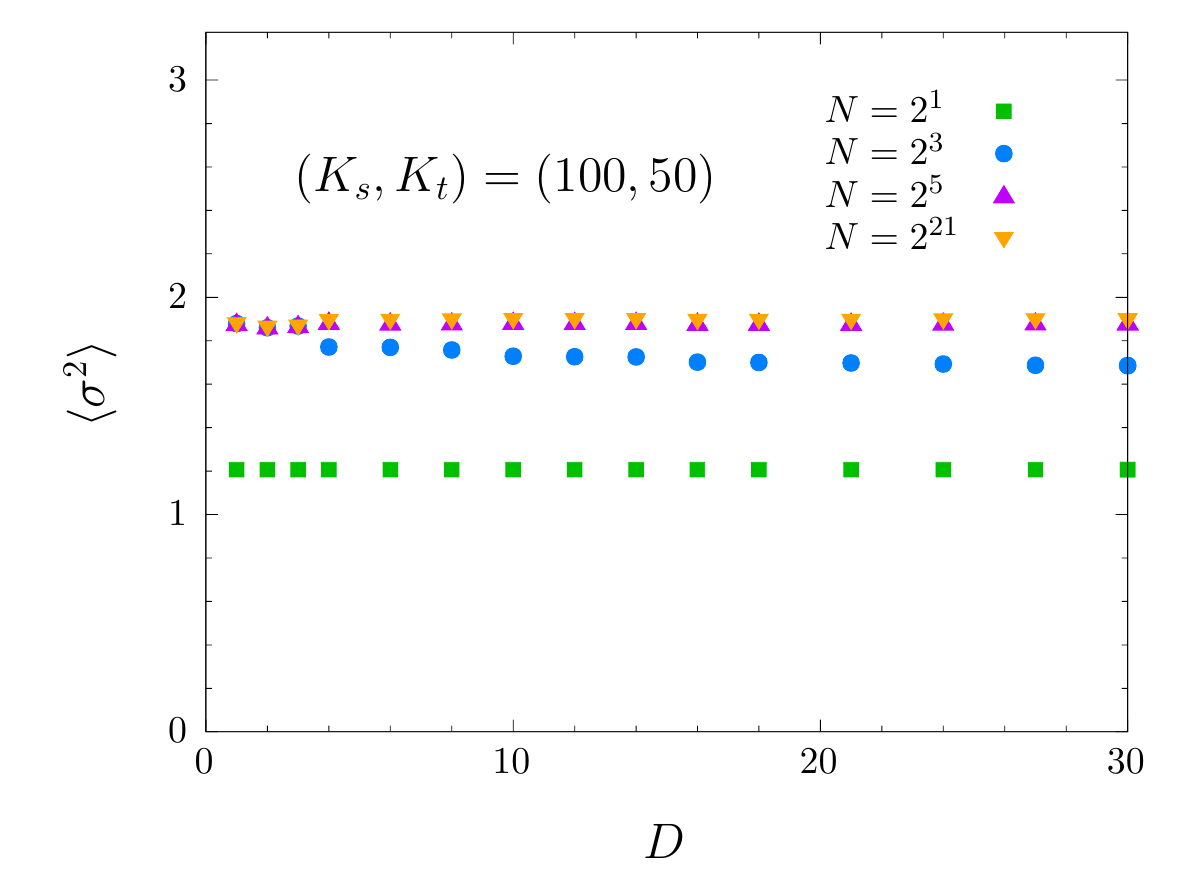}
  \end{center}
\end{minipage}
\caption{The $K_s, K_t$ and $D$ dependence of $\langle \sigma^2 \rangle$ at $\beta=1$, $N=2,2^3,2^5,2^{21}$ and $\mu=10^{-5}$.}
\label{fig:ls_B1}
\end{figure}

In Fig.~\ref{fig:s-mu}, the $\mu$ dependence of $\langle\sigma^2\rangle$ is shown for 
$N=2, 2^3, 2^{5}, 2^{21}$ at $\beta=0$ (Left) and $\beta=1$ (Right), 
where $K_s=100,K_t=50,D=30$ are fixed. 
At first glance, the result of $N = 2$ approaches a non zero finite value as $\mu$ decreases  
but in fact it does not converge with respect to $ K_t$. 
Figs.~\ref{fig:ls_B0} and \ref{fig:ls_B1} show $K_s, K_t $ and $D$ dependence of $\langle\sigma^2\rangle$ at $\mu=10^{-5}$. 
From these figures, we can confirm that the results of $N = 2$ are certainly not convergent. 
The analytic calculation given in Appendix \ref{sec:exact_results} tells us that $\langle\sigma^2\rangle=0$ for $N=2$.  
For large $N$, the results converge to non-zero finite values as $\mu\to0$. 
This confirms that the spike configuration is absent.

\begin{figure}[H]
\centering
\includegraphics[width=8cm]{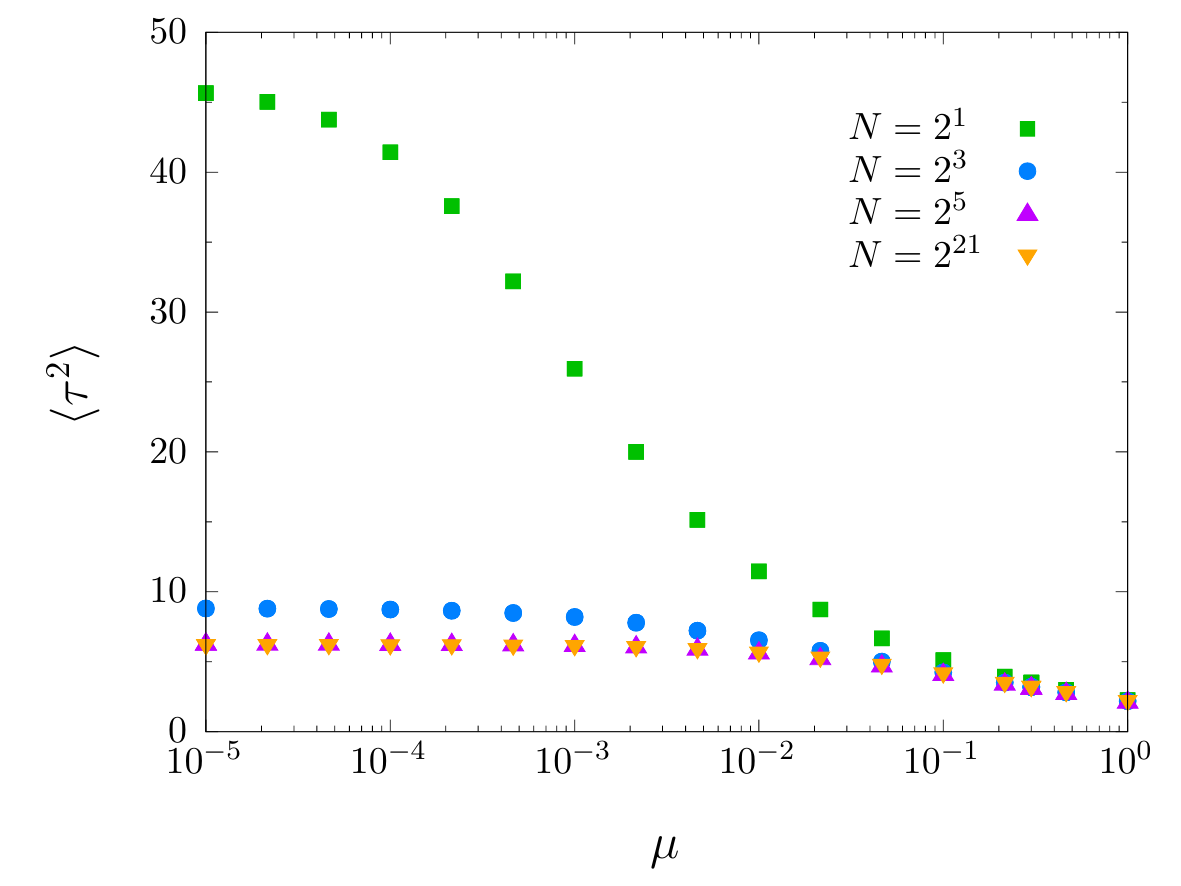}
\includegraphics[width=8cm]{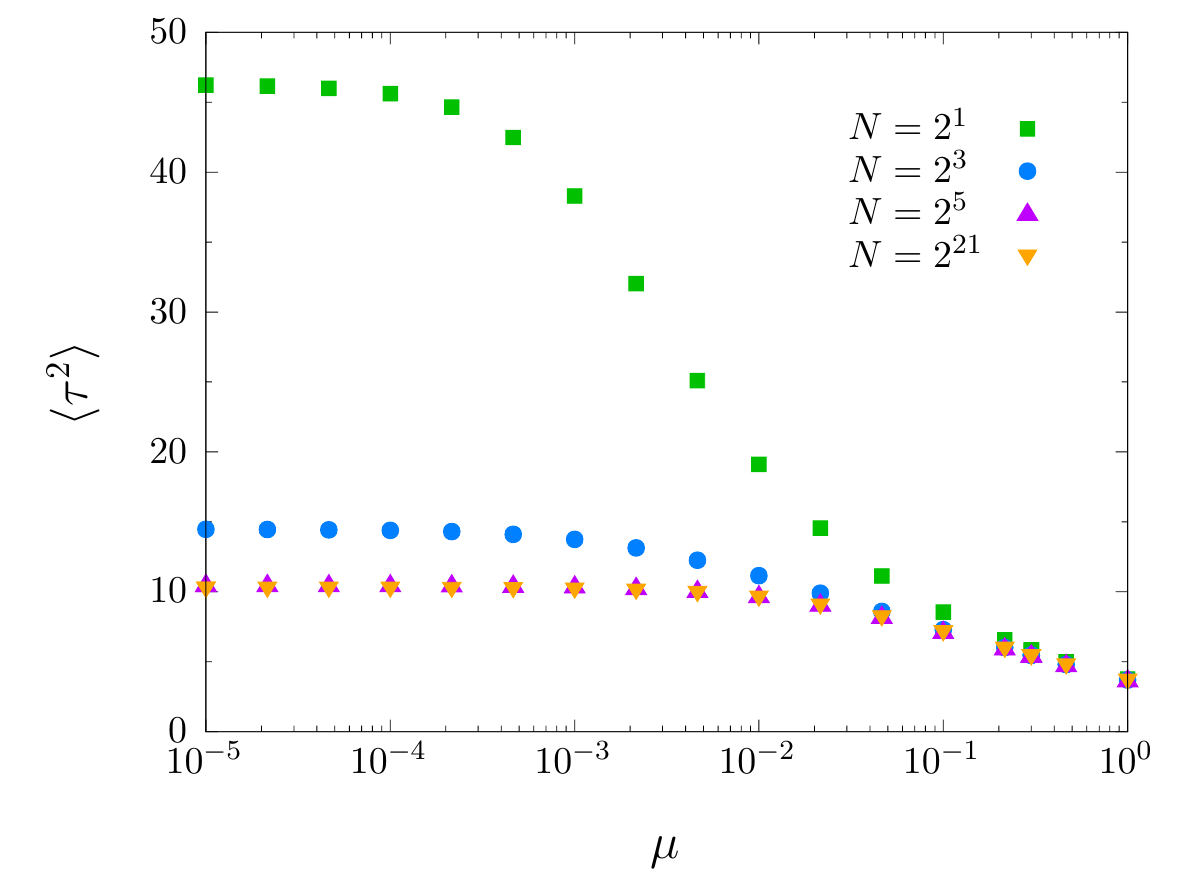}
\hspace{1cm}
\caption{The $\mu$ dependence of $\langle \tau^2\rangle$ for $\beta=0$ (Left) and $\beta=1$ (Right) at $N=2,2^3,2^5,2^{21}$, $(K_s,K_t)=(100,50)$ and $D=30$. 
At $N=2$, $\langle \tau^2\rangle$ increases as $\mu\to0$ and might diverge. 
The divergence or convergence as $\mu\to0$ becomes clear by investigating the convergence for $K_t,K_s$ and $D$ (Fig.~\ref{fig:lt_B0}, Fig.~\ref{fig:lt_B1}).}
\label{fig:t-mu}
\end{figure}

\begin{figure}[H]
\centering
\begin{minipage}{0.33\hsize}
  \begin{center}
   \includegraphics[width=5cm]{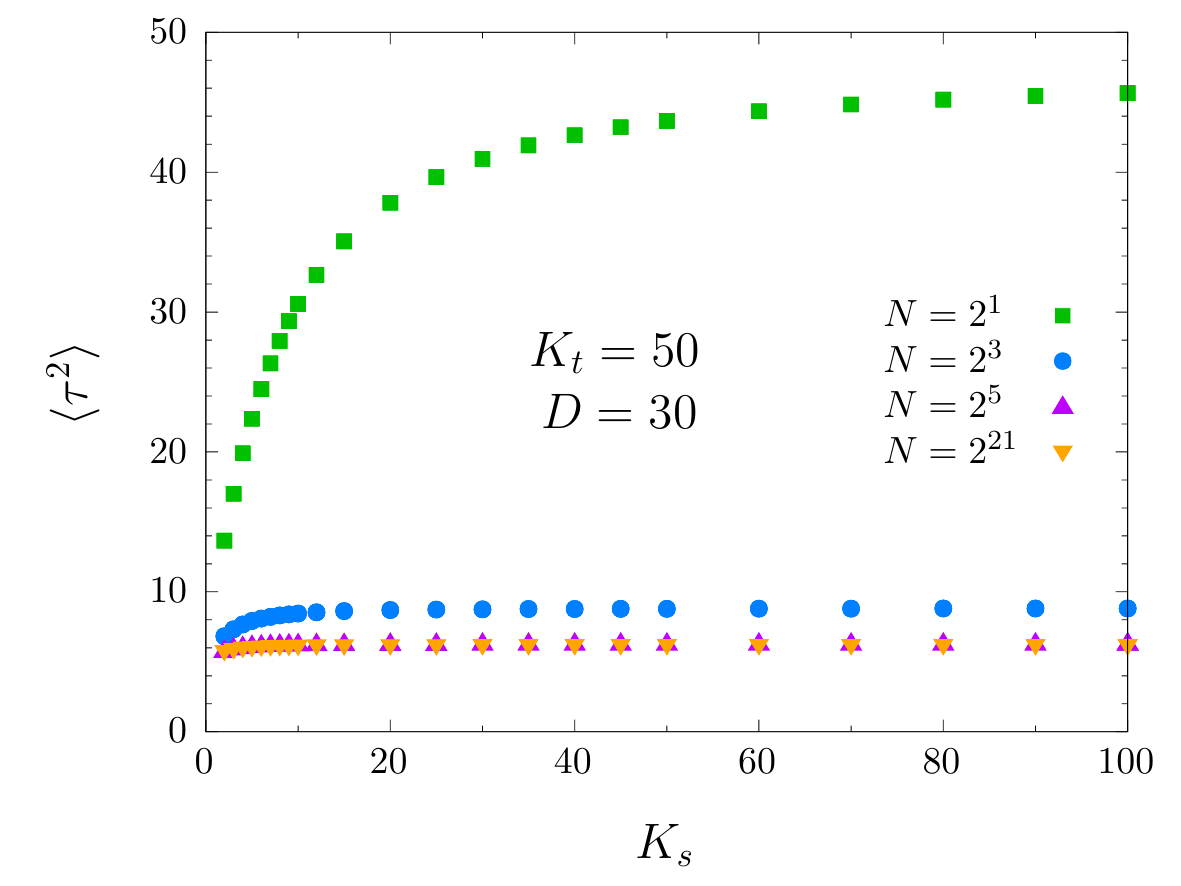}
  \end{center}
\end{minipage}%
\begin{minipage}{0.33\hsize}
  \begin{center}
   \includegraphics[width=5cm]{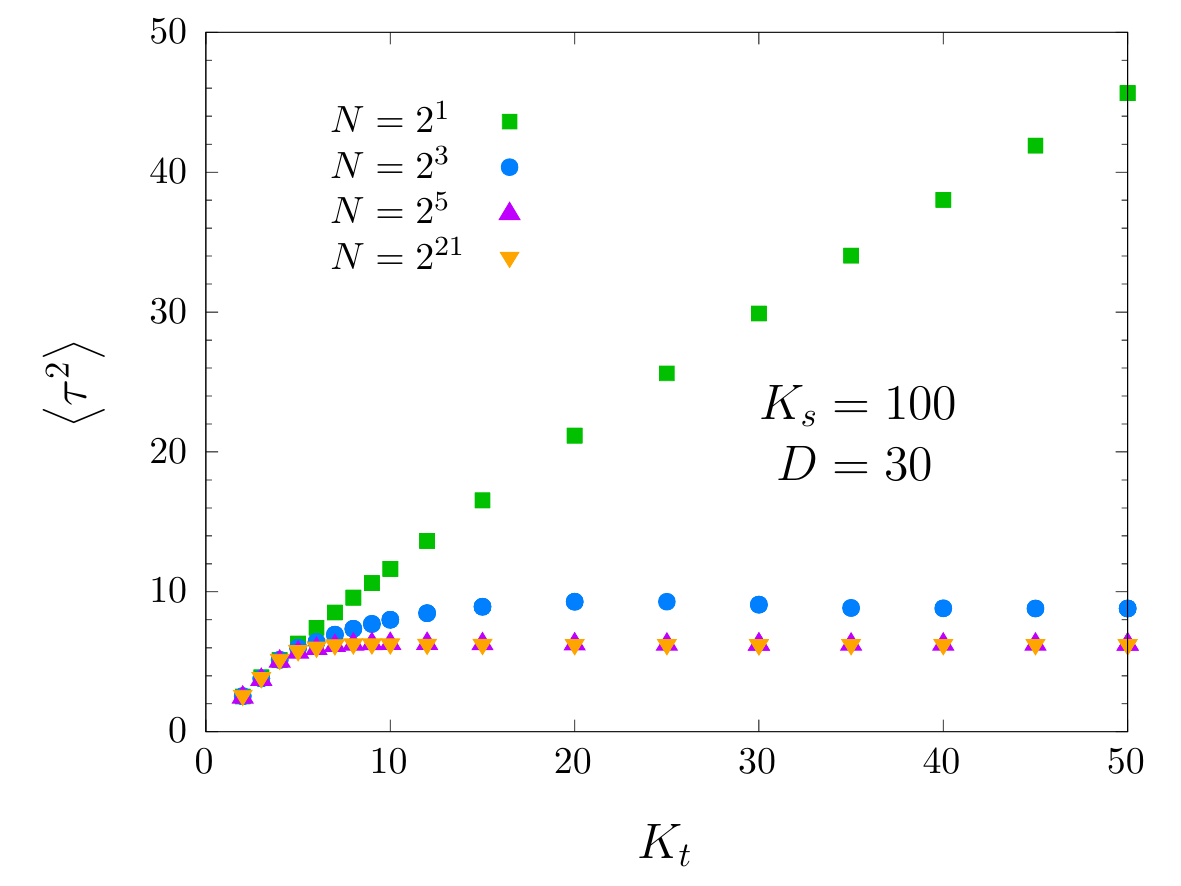}
  \end{center}
\end{minipage}%
\begin{minipage}{0.33\hsize}
  \begin{center}
   \includegraphics[width=5cm]{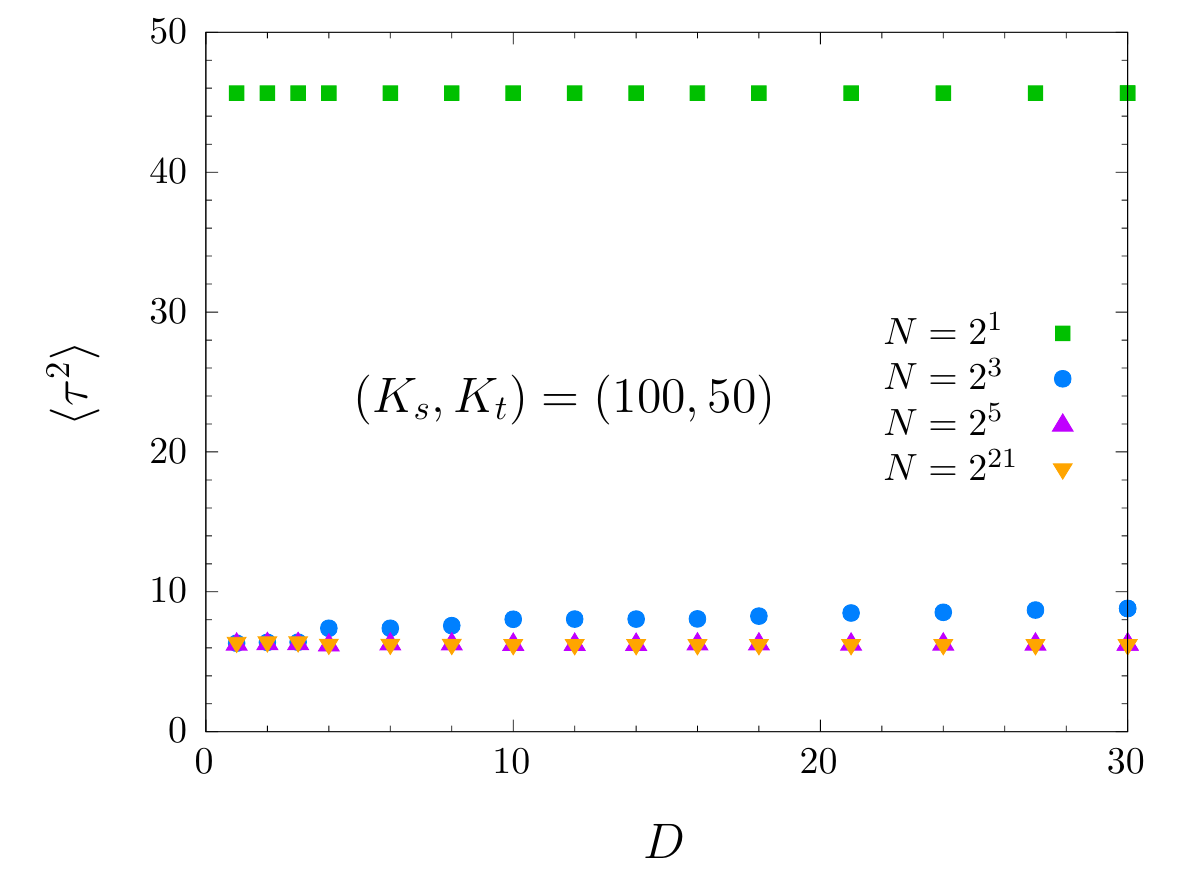}
  \end{center}
\end{minipage}
\caption{The $K_s,K_t$ and $D$ dependence of $\langle \tau^2 \rangle$ at $\beta=0$, $N=2,2^3,2^5,2^{21}$ and $\mu=10^{-5}$.}
\label{fig:lt_B0}
\end{figure}

\begin{figure}[H]
\centering
\begin{minipage}{0.33\hsize}
  \begin{center}
   \includegraphics[width=5cm]{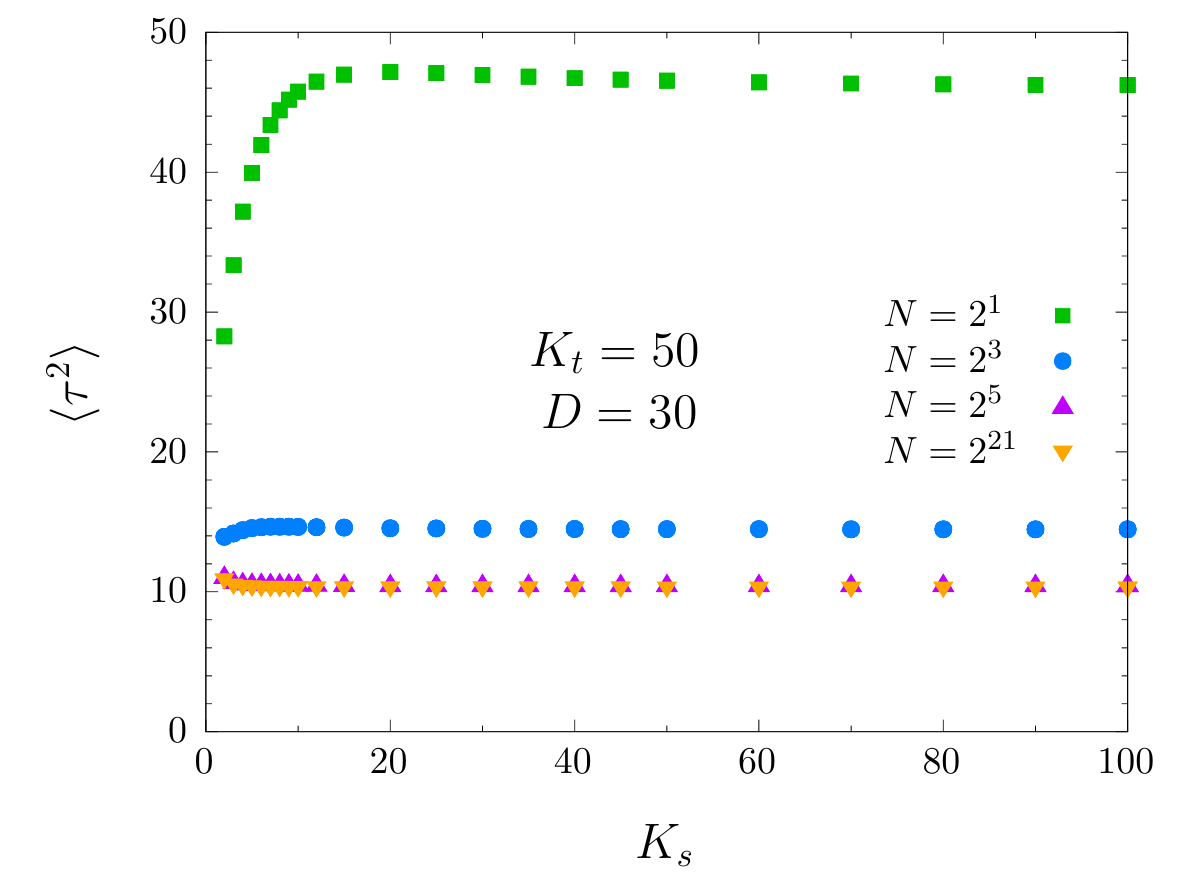}
  \end{center}
\end{minipage}%
\begin{minipage}{0.33\hsize}
  \begin{center}
   \includegraphics[width=5cm]{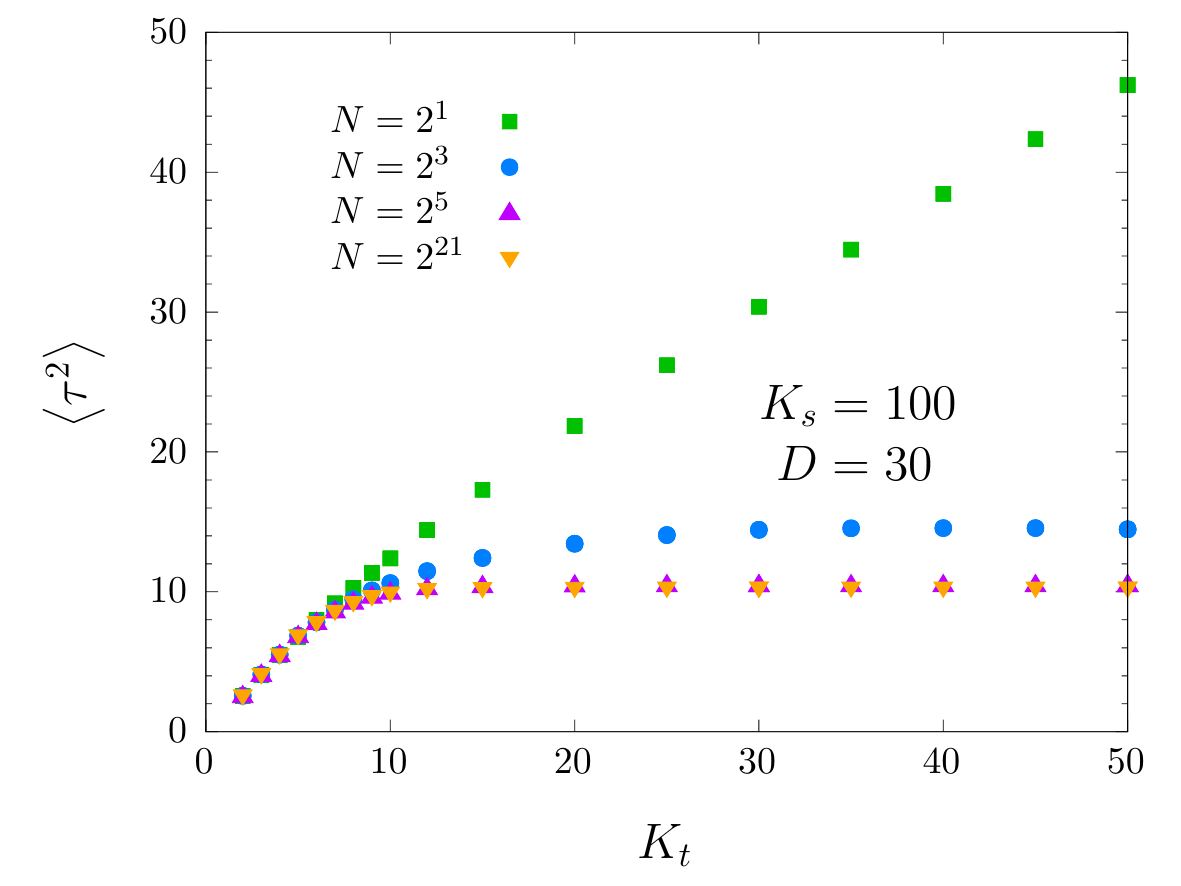}
  \end{center}
\end{minipage}%
\begin{minipage}{0.33\hsize}
  \begin{center}
   \includegraphics[width=5cm]{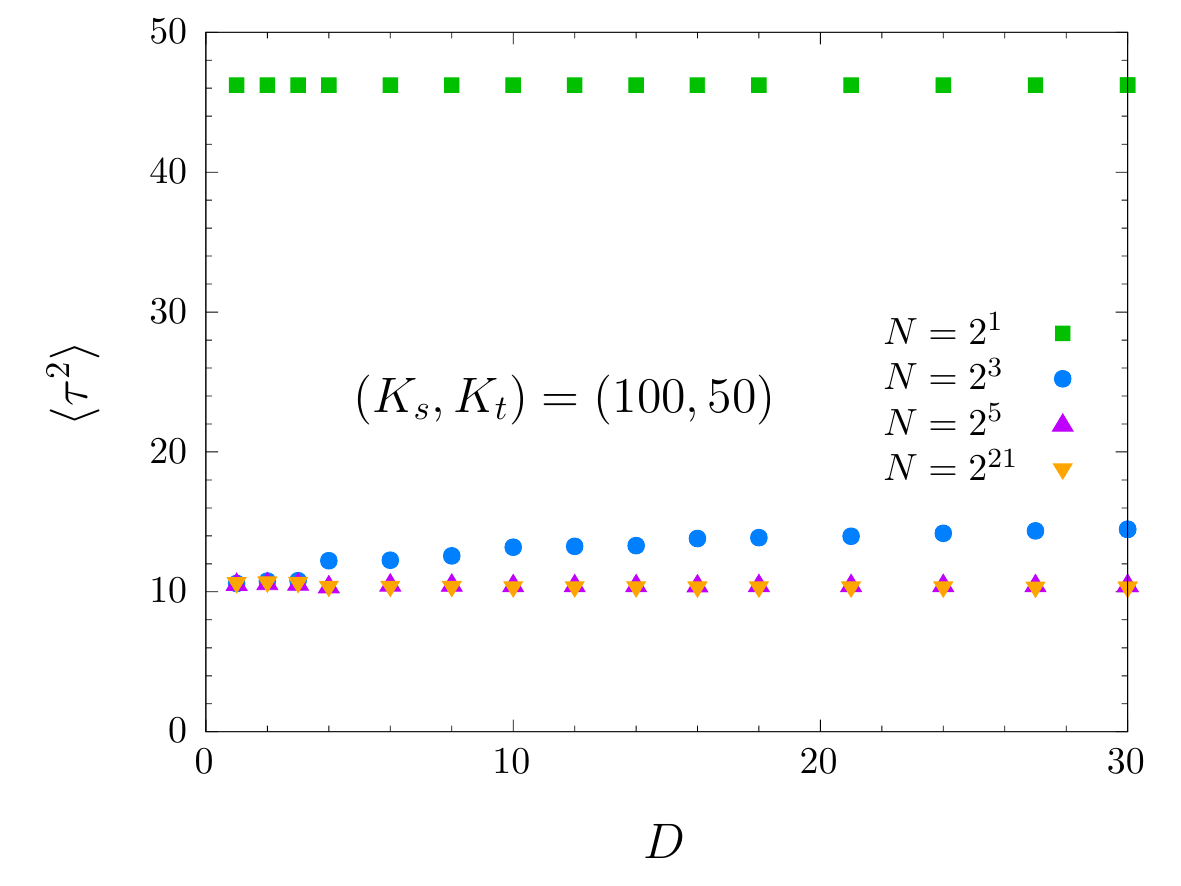}
  \end{center}
\end{minipage}
\caption{The $K_s,K_t$ and $D$ dependence of $\langle \tau^2 \rangle$ at $\beta=1$, $N=2,2^3,2^5,2^{21}$ and $\mu=10^{-5}$.}
\label{fig:lt_B1}
\end{figure}

Fig.~\ref{fig:t-mu} shows the $\mu$ dependence of $\langle\tau^2\rangle$ for 
$N=2, 2^3, 2^{5}, 2^{21}$ at $\beta=0$ (Left) and $\beta=1$ (Right), 
where $K_s=100$, $K_t=50$ and $D=30$ are fixed. 
For $N=2$, as expected from the analytic results shown in Appendix \ref{sec:exact_results}, 
$\langle\tau^2\rangle$ rapidly increases as $\mu$ decreases. 
The rate of increase is slowing down around $\mu\simeq 10^{-5}$.
This is because $K_s=100$, $K_t=50$ and $D=30$ are fixed. 
Figs.~\ref{fig:lt_B0} and \ref{fig:lt_B1} show the $K_s$ and $K_t,D$ dependence of $\langle\tau^2\rangle$ at $\mu=10^{-5}$.  
For $N=2$, $\langle\tau^2\rangle$ depends strongly on $K_t$.
In the large $K_t$ limit, the numerical value of  $\langle\tau^2\rangle$ at $\mu=10^{-5}$ for $N=2$ is much larger 
than those of Fig.~\ref{fig:t-mu}.  This implies that $\langle\tau^2\rangle$ diverges as $\mu \rightarrow 0$ 
as expected from the analytic result.

\begin{figure}[H]
\centering
\includegraphics[width=8cm]{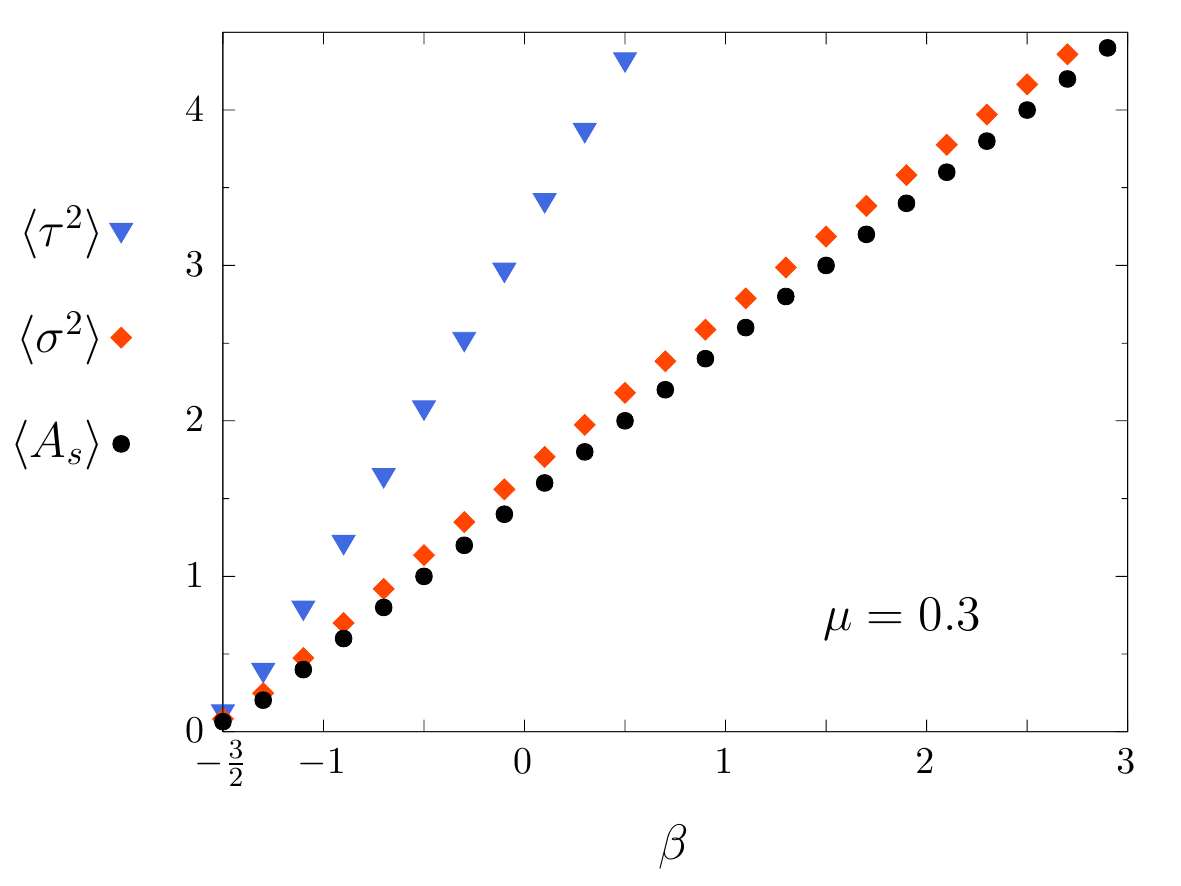}
\includegraphics[width=8cm]{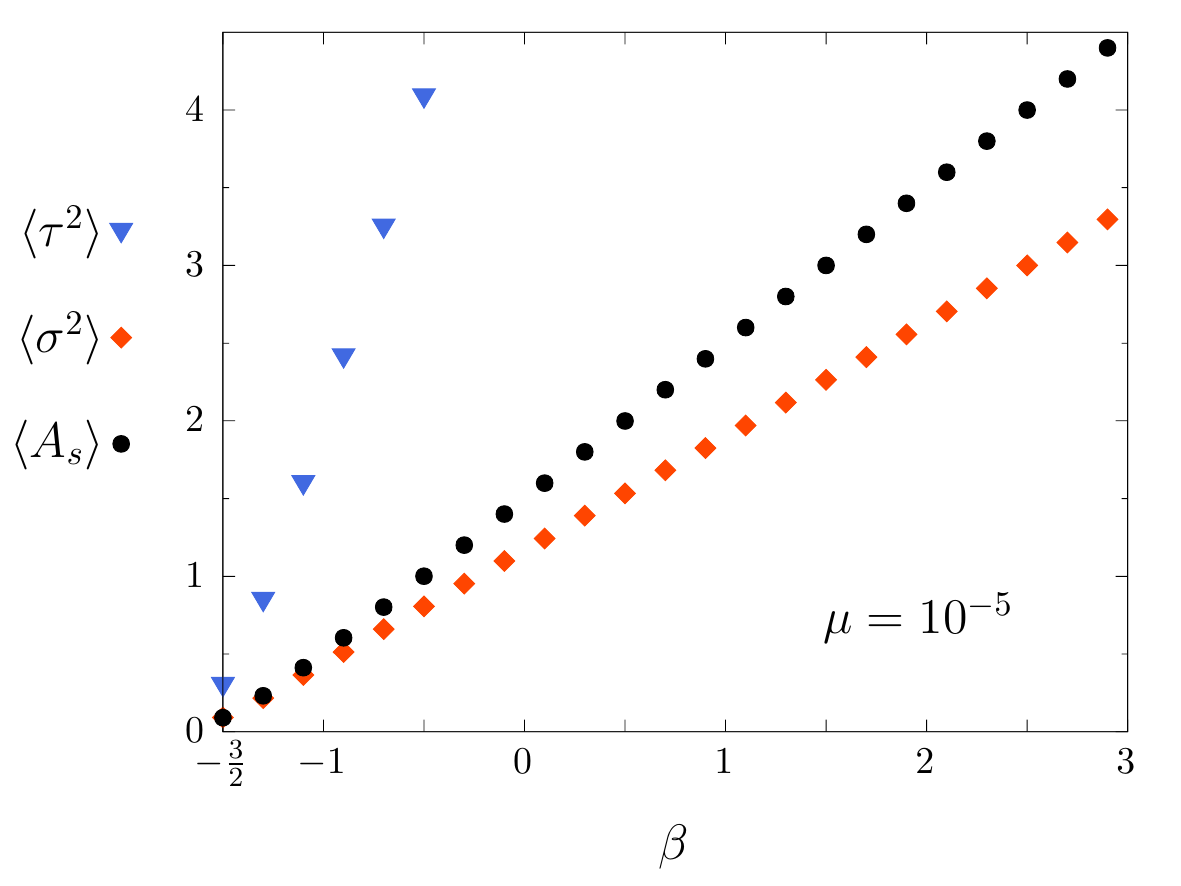}
\hspace{1cm}
\caption{The $\beta$ dependence of $\langle \tau^2 \rangle$, $\langle \sigma^2 \rangle$ and $\langle A_s \rangle$ for $\mu=0.3$ (Left) and $\mu=10^{-5}$ (Right) at $N=2^{21}$, $(K_s,K_t)=(100,50)$ and $D=30$. 
All of them converge to zero as $\beta\to-\frac{3}{2}$.}
\label{fig:A-t-s_beta}
\end{figure}

On the other hand, for $N=2^{21}$,   $\langle\tau^2\rangle$ converges to a finite value as $\mu$ decreases. 
As can be seen in Fig.~\ref{fig:lt_B0} and Fig.~\ref{fig:lt_B1}, the converged value is stable for $N=2^{21}$. 
From eq.~(\ref{eq:relation}), the fixed-area expectation value of $\tau^2$ is also finite.  
Fig.~\ref{fig:A-t-s_beta} shows the converged values of $\langle\sigma^2 \rangle$,  $\langle\tau^2\rangle$ and $\langle A_s \rangle$, 
for $N=2^{21}$. As $\beta$ approach $-3/2$, they smoothly approach zero.  
We thus conclude that $\langle\tau^2\rangle_A$ converges to the finite value as $\mu\to0$ in the limit 
where the  number of triangles is large. This means that the contribution of the pinched geometries might be suppressed in the limit, 
although to get more conclusive evidence it is quite important to check if the higher moments are also finite or not.



\section{Discussions}
\label{sec:Discussions}

We have constructed a tensor network representation of a $2$d Lorentzian model of QRC and 
demonstrated tensor renormalization group (TRG) calculations of the model. 
The  expectation value of space-time area reproduces the exact value in high accuracy. 
The Lorentzian model has a length divergent configuration called a pinched geometry.
Through the observation of the average edge lengths squared, 
we have deciphered a sign that the contribution of the pinched geometry would be suppressed in the limit 
where the number of simplices is large.

The reason why the pinched geometry would be suppressed is still unclear.  
When the number of triangles is two, the expectation value of any time-like edge length squared is divergent in numerical and analytic calculations. 
Our result implies that the contribution of the pinched geometries would be entropically suppressed when the number of triangles is large enough. 
However, the mechanism of this entropic suppression is not clear yet, and therefore this is completely a conjecture at this moment.  
Although the expectation values of squared edge lengths have been studied in this paper, 
we should investigate the case with various powers of edge lengths in the future further, 
in order to conclude if the pinched geometry is effectively absent in the $2$d Lorentzian QRC or not.  
Additionally, we have chosen a regular triangulation for convenience, and it is important to study other types of triangulation.   

The model explored in this paper by the TRG method does not have the sign problem thanks to the analytic continuation. 
This is due to the peculiarities of the model that is two-dimensional and has no matter fields. 
When coupling to matters or studying the higher dimensional models, the action cannot be real completely even if performing some analytic continuations. 
Since the tensor network approach effectively works for theories with the sign problem, 
the techniques established in this paper will be useful to handle Lorentzian QRC models in general.


\section*{Acknowledgement}
We would like to thank Jan Ambj\o{}rn,  Masafumi Fukuma, Masanori Hanada, Yusuke Namekawa, Jun Nishimura,
Naoki Sasakura 
and Tetsuyuki Yukawa   
for discussions and encouragements.   
This work was partially supported by JSPS KAKENHI Grant Number
19K03853,  19K14705, 	21K03537, 22H01222 and JST SPRING Grant Number JPMJSP2125. 
The computation was carried out using the supercomputer ``Flow'' at Information Technology Center, Nagoya University.


\appendix

\section{Analytic results for $N=2$}
\label{sec:exact_results}

The partition function \eqref{eq:partitionfun2} formally diverges due to the flat directions when $\mu=0$. 
In this appendix, we examine this divergence for $N=2$ from the exact calculation of the partition function. 

For $N=2$,  the partition function is given by
\[
Z = \int_0^\infty \ dx dy dz \ [A (x,y,z)]^{2\beta} \ e^{-2\lambda A (x,y,z) }\ , 
\]
where $A(x,y,z)$ is given by eq.~\eqref{area_def}. 
Changing the variables as $x=z r {\rm cos}{\theta}$ and $y=z r {\rm sin}{\theta}$ 
for $r \in [0,\infty)$ and $\theta \in [0,\pi/2]$ 
and integrating the $z$ variable, 
we have  
\[
Z = \frac{2^{3-2\beta}\Gamma(3+2\beta)}{\lambda^{3+2\beta}} \, \int_0^\infty dr  \int_0^{\pi/2} d \theta \,  \frac{r}{[f(r,\theta)]^\frac{3}{2}} \ ,
\label{z_n2}
\]
where 
\[
 f(r,\theta) =1 + 2 A(\theta)\, r +B(\theta, \mu) \, r^2   \ , 
\]
with $A(\theta)= {\rm cos}\theta + {\rm sin} \theta$ and $B(\theta,\mu)=1+\mu -{\rm sin} (2\theta)$.
At $\mu=0$, for large $r$, 
$f \sim r^2$ if $\theta \neq \pi/4$ 
while $f  \sim r$ if $\theta = \pi/4$. 
This implies $Z$ diverges when $\mu \rightarrow 0$.

We can integrate the $r$ variable of the partition function \eqref{z_n2}: 
\[
Z = \frac{2^{3-2\beta}\Gamma(3+2\beta)}{\lambda^{3+2\beta}} \,  \int_0^{\pi/2} d \theta \, g(\theta,\mu)\ , 
\]
where 
\[
g(\theta,\mu) = \frac{1}{A(\theta) \sqrt{B(\theta,\mu)} + B(\theta,\mu)}\ . 
\]
To investigate the divergence at $\theta=\pi/4$, let us change the variable as $\theta=\pi/4-s$. 
We then find that
\[
g(\pi/4-s, 0) = \frac{1}{2\vert s \vert} \quad \qquad {\rm for} \ s \ll 1\ , 
\]
and therefore $Z$ diverges logarithmically as $Z \sim {\rm log}(1/\mu)/2+{\cal O}(\mu^0)$ for $\mu\ll 1$.

The same calculation for $\langle \sigma^{2n} \rangle$ yields 
\[
\langle \sigma^{2n} \rangle = \left(\frac{2}{\lambda}\right)^n \frac{\Gamma(n+3+2\beta)}{\Gamma(3+2\beta) }\, 
\frac{\int_0^\infty dr  \int_0^{\pi/2} d \theta \,  r \, [f(r,\theta)]^{-(n+3)/2}}
{\int_0^\infty dr  \int_0^{\pi/2} d \theta \,  r \, [f(r,\theta)]^{-3/2} }\ . 
\label{sigma2n}
\]
Note that the large $r$ behavior of $r [f(r,\theta)]^{-(n+3)/2}$
is milder than $r \, [f(r,\theta)]^{-3/2}$ for $n>1$. 
We then obtain 
\[
\lim_{\mu \rightarrow 0}
\langle \sigma^{2n} \rangle=0\ , 
\]
since the integral of the denominator diverges at least faster than that of the numerator.

Concerning $\langle \tau_1^{2n}  \rangle$, using the same techniques of integration, 
we obtain
\[
\langle \tau_1^{2n} \rangle = \left(\frac{2}{\lambda}\right)^n \frac{\Gamma(n+3+2\beta)}{\Gamma(3+2\beta) }\, 
\frac{\int_0^\infty dr  \int_0^{\pi/2} d \theta \,  r^{n+1} \,h(\theta)\,  [f(r,\theta)]^{-(n+3)/2}}
{\int_0^\infty dr  \int_0^{\pi/2} d \theta \,  r \, [f(r,\theta)]^{-3/2} }\ ,
\label{tau2n}
\]
where $h(\theta)={\rm cos}^n \theta$.
The large $r$ behavior of the integrand of the numerator is different from that for $\langle \sigma^{2n}  \rangle$.
At $\theta=\pi/4$, we find that $r^{n+1} [f(r,\theta)]^{-(n+3)/2} \sim r^{(n-1)/2}$, 
which is larger than that of the denominator for $n>1$. 
We thus obtain 
\[
\lim_{\mu \rightarrow 0}
\langle \tau_1^{2n} \rangle= \infty\ , 
\]
since the integral of the numerator diverges faster than that of the denominator. 
The expectation value of $\tau_1^{m} \tau_2^{n-m}$ ($m=0,1,\cdots,n$) also diverges 
because the value of $h(\theta)={\rm cos}^m \theta \, {\rm sin}^{n-m} \theta$ at $\theta=\pi/4$ does not change.

\section{Analytic continuation}
\label{sec:AnalyticContinuation}

We consider the regular triangulation shown in Fig.~\ref{fig:triangulation} with the periodic boundary condition for both space-like and time-like directions. 
$N$ and $N_e$ are the numbers of triangles and edges where $N$ is an even integer and 
$N_e=3N/2$. 
The Lorentzian partition function we consider is given by  
\[
Z
= \int [d\ell^2] \
e^{i (\lambda+ i \epsilon) \sum^N_{s=1}A (\{\ell^2\})
}\ ,  
\label{eq:partitionfun3}
\]
where the integral measure is given by (\ref{eq:measure}); 
$\lambda \neq 0$, $\mu>0$ and $\epsilon>0$; 
we assume that $\beta>-3/2$. 
Here we suppressed the index $\mu$ of $A^{(\mu)}$ for notational simplicity.

We first focus on the case of $\lambda>0$.   
Changing $N_e$ variables $\{\ell^2_e\}$ into the $n$-dimensional spherical coordinates $(r,\Omega_{n-1})=(r,\phi_1, \cdots, \phi_{n-1})$
as done in \cite{Tate:2011ct},
we have
\[
Z
= \int^{\infty}_{0} dr \ r^{N(\beta + 3/2)-1} \ \int d\Omega_{n-1} 
G_{\beta}(\Omega_{n-1})
e^{i (\lambda + i\epsilon ) r F (\Omega_{n-1})}\ , 
\label{eq:partitionfun4}
\]
where 
\[
&[d\ell^2] = \vcentcolon  r^{N(\beta+3/2)-1} G_\beta(\Omega_{n-1}) dr d \Omega_{n-1}\ , \\
&\sum^N_{s=1}A (\{\ell^2\}) = \vcentcolon r F(\Omega_{n-1})\ . 
\] 
Note that $F>0$ for any $\mu>0$.

\begin{figure}[h]
\centering
\includegraphics[width=2in]{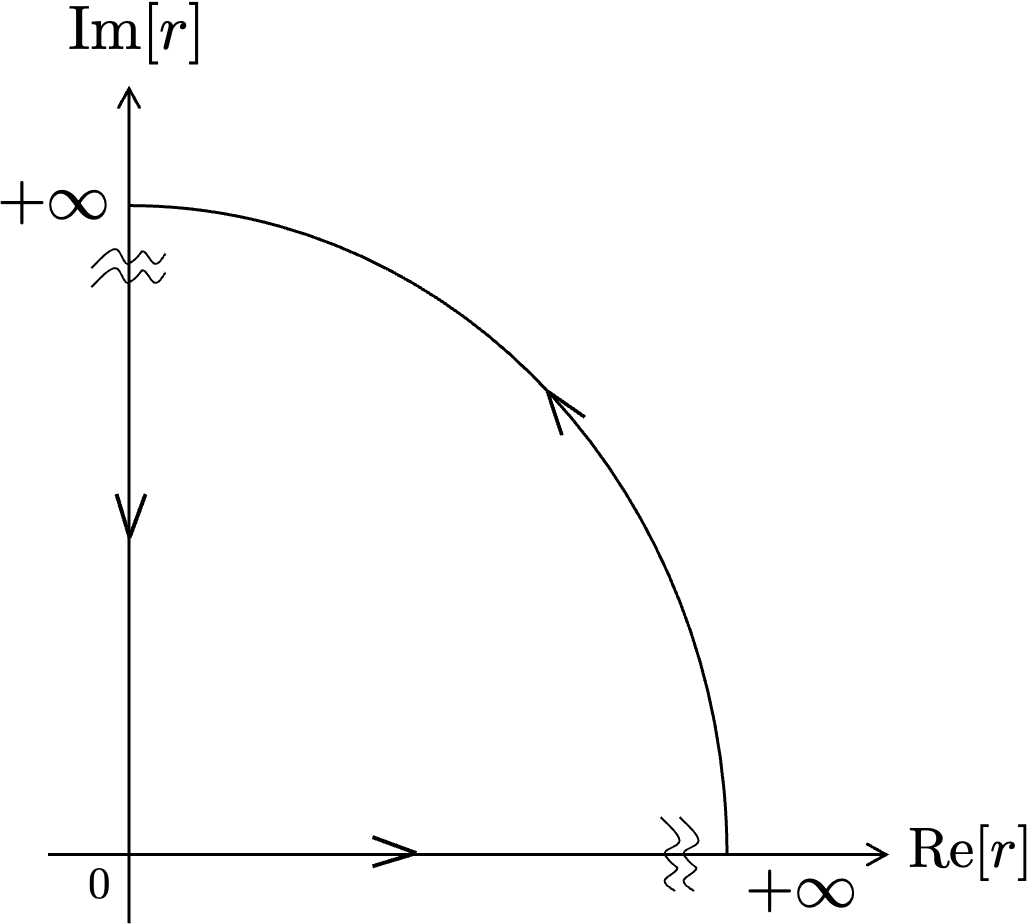}
\caption{Change of the integration contour.}
\label{fig:path}
\end{figure}

Let us first consider the positive $\lambda$ case and change the integration contour for the $r$-integral from the real axis $(0,\infty)$ to the imaginary axis $(0,i \infty)$
by choosing the contour as the one shown in Fig.~\ref{fig:path}.
The partition function can be expressed as
\[
Z 
&= i^{N(\beta+3/2)} 
\int^{\infty}_{0} dr \ r^{N(\beta + 3/2)-1} \ \int d\Omega_{n-1} 
G_{\beta}(\Omega_{n-1})
e^{-(\lambda + i\epsilon ) r F (\Omega_{n-1})}\ , 
\label{eq:partitionfun5}
\]
because the integrand for $r=Re^{i \theta}$ for $0<\theta<\pi/2$ is rapidly damping as $R \rightarrow \infty$.  
The limit of $\epsilon \rightarrow 0$ can be safely taken for eq.~\eqref{eq:partitionfun5}. 
Thus the partition function of the Lorentzian QRC is defined by eq.~\eqref{eq:partitionfun2}.
Concerning the case of $\lambda<0$, we essentially follow the same step and in particular change the integration contour for the $r$-integral from the real axis $(0,\infty)$ to the imaginary one $(0,-i \infty)$. 
%



\end{document}